%% file: NonclassicalStates.tex
\documentclass[vecphys]{svmult}

\usepackage{amsmath}
\usepackage{amssymb}

\usepackage{mathptmx}       
\usepackage{helvet}         
\usepackage{courier}        
\usepackage{makeidx}         
\usepackage{graphicx}        
\usepackage{multicol}        
\usepackage[bottom]{footmisc}

\makeindex             

\newcommand{\OmegaM}{\Omega_{{\rm M}}}

\newcommand{\G}{G}

\newcommand{\omegaopt}{\omega_{{\rm opt}}}

\newcommand{\mass}{m_{{\rm eff}}}

\newcommand{\bh}{\hat{b}}

\newcommand{\ah}{\hat{a}}

\newcommand{\nth}{\bar{n}_{{\rm th}}}

\newcommand{\gom}{g}

\newcommand{\gomzero}{g_{0}}

\newcommand{\GammaM}{\Gamma_{{\rm M}}}

\newcommand{\GammaEff}{\Gamma_{\rm eff}}

\newcommand{\cavlength}{L}

\newcommand{\finesse}{\mathcal{F}}

\newcommand{\Pin}{P_{{\rm in}}}

\newcommand{\Langevin}{\xi}

\newcommand{\intdb}[4]{\ensuremath{\int_{#2}^{#3}\!\!\mathrm{d}#4\,}}

\begin{document}

\title*{Nonclassical States of Light and Mechanics}

\author{Klemens Hammerer, Claudiu Genes, David Vitali, Paolo Tombesi, Gerard Milburn, Christoph Simon, Dirk Bouwmeester}

\institute{Klemens Hammerer \at Leibniz University Hanover, \email{Klemens.Hammerer@itp.uni-hannover.de}
\and Claudiu Genes \at Univerity of Innsbruck, \email{Claudiu.Genes@uibk.ac.at}
\and David Vitali \at University of Camerino, \email{David.Vitali@unicam.it}
\and Paolo Tombesi \at University of Camerino, \email{Paolo.Tombesi@unicam.it}
\and Gerard Milburn \at University of Queensland, \email{Milburn@physics.uq.edu.au}
\and Christoph Simon \at University of Calgary, \email{}
\and Dirk Bouwmeester \at University of California, \email{Bouwmeester@physics.ucsb.edu}
}

\maketitle
\abstract{This chapter reports on theoretical protocols for generating nonclassical states of light and mechanics. Nonclassical states are understood as squeezed states, entangled states or states with negative Wigner function, and the nonclassicality can refer either to light, to mechanics, or to both, light and mechanics. In all protocols nonclassicallity arises from a strong optomechanical coupling. Some protocols rely in addition on homodyne detection or photon counting of light.}

\section{Introduction}

\input{Part0_Hammerer}


\section{Non-classical states of light}\label{Sec:Light}

\input{Part1_Vitali-Tombesi}

\input{Part2_Genes}


\section{Non-classical states of mechanics}\label{Sec:Mechanics}

\input{Part3_Genes}

\input{Part4_Milburn}

\subsection{Non-Gaussian state via interaction with single photons and photon counting}

\input{Part5_Bouwmeester}


\section{Entangled states of mechanics and light}\label{Sec:Light&Mechanics}

\input{Part6_Genes}

\subsection{Entanglement with pulsed light}

\input{Part7_Hammerer}

\section{Conclusion and Outlook}

\input{Part8_Hammerer}

\bibliographystyle{spphys}
\bibliography{Nonclassical}

\end{document}

%% file: Part0_Hammerer.tex
An outstanding goal in the field of optomechanics is to go beyond the regime of classical physics, and to generate nonclassical states, either in light, the mechanical oscillator, or involving both systems, mechanics and light. The states in which light and mechanical oscillators are found naturally are those with Gaussian statistics with respect to measurements of position and momentum (or field quadratures in the case of light). The class of Gaussian states include for example thermal states of the mechanical mode, as well as its ground state, and on the side of light coherent states and vacuum. These are the sort of classical states in which optomechanical systems can be prepared easily. In this chapter we summarize and review means to go beyond this class of states, and to prepare \textit{nonclassical} states of optomechanical systems.

Within the family of Gaussian states those states are usually referred to as nonclassical in which the variance of at least one of the canonical variables is reduced below the noise level of zero point fluctuations. In the case of a single mode, \textit{e.g.} light or mechanics, these are \textit{squeezed states}. If we are concerned with a system comprised of several modes, \textit{e.g.} light and mechanics or two mechanical modes, the noise reduction can also pertain to a variance of a generalized canonical variable involving dynamical degrees of freedom of more than one mode. Squeezing of such a collective variable can arise in a state bearing sufficiently strong correlations among its constituent systems. For Gaussian states it is in fact true that this sort of squeezing provides a necessary and sufficient condition for the two systems to be in an inseparable, quantum mechanically \textit{entangled state}. Nonclassicality within the domain of Gaussian states thus means to prepare squeezed or entangled states.

For states exhibiting non Gaussian statistics the notion of nonclassicality is less clear. One generally accepted criterium is based on the Wigner phase space distribution. A state is thereby classified as non classical when its Wigner function is non positive. This notion of nonclassicality in fact implies for pure quantum states that all non Gaussian states are also non classical since every pure non Gaussian quantum state has a non positive Wigner function. For mixed states the same is not true. Under realistic conditions the state of optomechanical systems will necessarily be a statistical mixture such that the preparation and verification of states with a non positive Wigner distribution poses a formidable challenge. Paradigmatic states of this kind will be states which are close to eigenenergy (Fock) states of the mechanical system.

Optomechanical systems present a promising and versatile platform for creation and verification of either sort of nonclassical states. Squeezed and entangled Gaussian states are in principle achievable with the strong, linearized form of the radiation pressure interaction, or might be conditionally prepared and verified by means of homodyne detection of light. These are all ``Gaussian tools'' which conserve the Gaussian character of the overall state, but are sufficient to steer the system towards Gaussian non classical states. In order to prepare non Gaussian states, possibly with negative Wigner function, the toolbox has to be enlarged in order to encompass also some non Gaussian instrument. This can be achieved either by driving the optomechanical system with a non Gaussian state of light, such as a single photon state, or by preparing states conditioned on a photon counting event. Ultimately the radiation pressure interaction itself is a nonlinear interaction (cubic in annihilation/creation operators) and therefore does in principle generate non Gausssian states for sufficiently strong coupling $\gomzero$ at the single photon level. Quite generally one can state that some sort of strong coupling condition has to be fulfilled in any protocol for achieving a nonclassical state. Fulfilling the respective strong coupling condition is thus the experimental challenge on the route towards nonclassicality in optomechanics.

In the following we will present a selection of strategies aiming at the preparation of nonclassical states. In Sec.~\ref{Sec:Light} we review ideas of using an optomechanical cavity as source of squeezed and entangled light. Central to this approach is the fact that the radiation pressure provides an effective Kerr nonlinearity for the cavity, which is well known to be able to generate squeezing of light. In Sec.~\ref{Sec:Mechanics} we discuss nonclassical states of the mechanical mode. This involves e.g. the preparation of squeezed states as well as non Gaussian states via state transfer form light, continuous measurement in a nonlinearly coupled optomechanical system, or interaction with single photons and photon counting. Sec.~\ref{Sec:Light&Mechanics} is devoted to nonclassical states involving both systems, light and mechanics, and summarizes ideas to prepare the optomechanical system in an entangled states, either in steady state under continuous wave driving fields, or via interaction with pulsed light.

%% file: Part1_Vitali-Tombesi.tex
\subsection{Ponderomotive squeezing}\label{sec:1}

One of the first predictions of quantum effects in cavity optomechanical system concerned ponderomotive squeezing \cite{Mancini1994, Fabre1994}, i.e., the possibility to generate quadrature-squeezed light at the cavity output due to the radiation pressure interaction of the cavity mode with a vibrating resonator. The mechanical element is shifted proportionally to the intracavity intensity, and consequently the optical path inside the cavity depends upon such intensity. Therefore the optomechanical system behaves similarly to a cavity filled with a nonlinear Kerr medium. This can be seen also by inserting the formal solution of the time evolution of the mechanical displacement $\hat{x}(t)$ into the Quantum Langevin equation (QLE) for the cavity field annihilation operator $\ah (t)$,
\begin{equation}\label{eq:formalsol}
    \dot{\ah}=-\left[\frac{\kappa}{2}+i\omegaopt(0)\right]\ah + \int_{-\infty}^t ds \chi_M(t-s)\left[i\hbar \G^2\ah(t) \ah^{\dagger}(s) \ah(s)+i\G \ah(t)\Langevin (s)\right]+\sqrt{\kappa}a_{in}(t),
\end{equation}
where $a_{in}(t)$ is the driving field (including the vacuum field) and
\begin{equation}\label{eq:susc}
   \chi_M(t)=\int_{-\infty}^{\infty} \frac{d\omega}{2\pi}\frac{e^{-i \omega t}}{\mass \left(\OmegaM^2-\omega^2-i \GammaM \omega\right)}=\frac{e^{-\GammaM t/2}}{\mass \tilde{\Omega}_M}\sin \tilde{\Omega}_M t
\end{equation}
is the mechanical susceptibility (here $\tilde{\Omega}_M=\sqrt{\OmegaM ^2-\GammaM^2/4}$). Eq.~(\ref{eq:formalsol}) shows that the optomechanical coupling acts as a Kerr nonlinearity on the cavity field, but with two important differences: i) the effective nonlinearity is delayed by a time depending upon the dynamics of the mechanical element; ii) the optomechanical interaction transmits mechanical thermal noise $\Langevin (t)$ to the cavity field, causing fluctuations of its frequency. When the mechanical oscillator is fast enough, i.e., we look at low frequencies $\omega \ll \OmegaM $, the mechanical response is instantaneous, $ \chi_M(t) \simeq \delta(t)/\mass \OmegaM^2$, and the nonlinear term becomes indistinguishable from a Kerr term, with an effective nonlinear coefficient $\chi^{(3)}=\hbar \G^2/\mass \OmegaM^2$.

It is known that when a cavity containing a Kerr medium is driven by an intense laser, one gets appreciable squeezing
in the spectrum of quadrature fluctuations at the cavity output \cite{Walls1995}. The above analogy therefore suggests that a strongly driven optomechanical cavity will also be able to produce quadrature squeezing at its output, provided that optomechanical coupling predominates over the detrimental effect of thermal noise \cite{Mancini1994,Fabre1994}.

We show this fact by starting from the Fourier-transformed linearized QLE for the fluctuations around the classical steady state
\begin{eqnarray}
\mass \left(\OmegaM ^2-\omega^2-i \omega \GammaM\right) x(\omega)   &  = &
\hbar \G \alpha_s \delta X(\omega) +\Langevin(\omega), \label{QLElinearft2} \\
\left(\frac{\kappa}{2}-i \omega \right)\delta X(\omega)  &  = & -\Delta\delta Y(\omega)+\sqrt{\kappa}\delta X^{in}(\omega),\label{QLElinearft3} \\
\left(\frac{\kappa}{2}-i \omega \right) \delta Y(\omega)  &  = & \Delta\delta X(\omega)+\G \alpha_s x(\omega)+\sqrt{\kappa}\delta Y^{in}(\omega), \label{QLElinearft4}
\end{eqnarray}
where we have chosen the phase reference so that the stationary amplitude of the intracavity field $\alpha_s$ is real, $\delta X=\delta \ah+\delta \ah^{\dagger}$ [$\delta Y=-i\left(\delta \ah-\delta \ah^{\dagger}\right)$] is the amplitude (phase) quadrature of the field fluctuations, and $\delta X^{in}$  and $\delta Y^{in}$ are the corresponding quadratures of the vacuum input field.
The output quadrature noise spectra are obtained solving Eqs.~(\ref{QLElinearft2})-(\ref{QLElinearft4}), and by using input-output relations \cite{Walls1995},
the vacuum input noise spectra $S_X^{in}(\omega)=S_Y^{in}(\omega)=1$, and the fluctuation-dissipation theorem for the thermal spectrum $S_{\Langevin}(\omega)=\hbar \omega \GammaM \mass \coth \left(\hbar\omega /2k_B T\right)$.

The output light is squeezed at phase $\phi$ when the corresponding noise spectrum is below the shot-noise limit, $S_{\phi}^{out}(\omega)<1$, where
$S_{\phi}^{out}(\omega)= S_{X}^{out}(\omega)\cos^2\phi+S_{Y}^{out}(\omega)\sin^2 \phi+S_{XY}^{out}(\omega)\sin 2\phi $,
and the amplitude and phase noise spectra $S_{X}^{out}(\omega)$ and $S_{y}^{out}(\omega)$ satisfy the Heisenberg uncertainty theorem $S_{X}^{out}(\omega)S_{Y}^{out}(\omega) > 1+\left[S_{XY}^{out}(\omega)\right]^2$ \cite{Braginsky1995}.
However, rather than looking at the noise spectrum at a fixed phase of the field, one usually performs an optimization and considers, for every frequency $\omega$, the field phase $\phi_{opt}(\omega)$ possessing the minimum noise spectrum, defining in this way the \emph{optimal squeezing spectrum},
\begin{eqnarray}\label{eq:speopt}
    S_{opt}(\omega)&=&\min_{\phi}S_{\phi}^{out}(\omega)\nonumber\\
    &=&\frac{2S_{X}^{out}(\omega)S_{Y}^{out}(\omega)-2\left[S_{XY}^{out}(\omega)\right]^2}
    {S_{X}^{out}(\omega)+S_{Y}^{out}(\omega) +\sqrt{\left[S_{X}^{out}(\omega)  -S_{Y}^{out}(\omega)\right]^2+ 4\left[S_{XY}^{out}(\omega)\right]^2}}.
\end{eqnarray}
The frequency-dependent optimal phase is correspondingly given by
\begin{equation}\label{eq:phiopt}
   \phi_{opt}(\omega)=\frac{1}{2}\arctan\left[\frac{2S_{XY}^{out}(\omega)}{S_{X}^{out}(\omega)-S_{Y}^{out}(\omega)}\right].
\end{equation}
Since there is no benefit for choosing a nonzero detuning, we restrict to the resonant case $\Delta = 0$, which is always stable and where expressions are  simpler. One gets
\begin{eqnarray}
  S_X^{out}(\omega) &=&  1,\;\;\;\; \;\;\;S_{XY}^{out}(\omega)=\frac{\kappa \hbar \G^2\alpha_s^2 {\rm Re}\left\{\chi_M(\omega)\right\}}{\kappa^2/4+\omega^2},\label{eq:sxy}\\
 S_{Y}^{out}(\omega)&=& 1+S_{XY}^{out}(\omega)^2+S_{r}(\omega), \label{eq:sy}
\end{eqnarray}
where
\begin{equation}\label{eq:speres}
  S_{r}(\omega)=  \left[\frac{\kappa \hbar \G^2\alpha_s^2 {\rm Im}\left\{\chi_M(\omega)\right\}}{\kappa^2/4+\omega^2}\right]^2 +\frac{\kappa \hbar \G^2\alpha_s^2 {\rm Im}\left\{\chi_M(\omega)\right\}}{\kappa^2/4+\omega^2} \coth \left(\frac{\hbar \omega}{2k_B T}\right).
\end{equation}
Inserting Eqs.~(\ref{eq:sxy})-(\ref{eq:sy}) into Eq.~(\ref{eq:speopt}) one sees that the strongest squeezing is obtained when the two limits $S_{r}(\omega) \ll 1$ and $\left[S_{XY}^{out}(\omega)\right]^2 \gg 1$ are simultaneously satisfied. These conditions are already suggested by Eq.~(\ref{eq:formalsol}): $S_{r}(\omega) \ll 1$ means that thermal noise is negligible, which occurs at low temperatures and small mechanical damping ${\rm Im}\left\{\chi_M(\omega)\right\}$, i.e., large mechanical quality factor $Q$; $\left[S_{XY}^{out}(\omega)\right]^2 \gg 1$ means large radiation pressure, achieved at large intracavity field and small mass. Ponderomotive squeezing is therefore attained when
\begin{equation}
\frac{[S_{XY}^{out}(\omega)]^2}{S_r(\omega)} \sim \frac{\Pin\omega_{L}}{\mass c^2 \OmegaM^2}\frac{\finesse^2 Q}{\nth} \gg 1,
\end{equation}
and in this ideal limit 
$ S_{opt}(\omega) \simeq \left[S_{XY}^{out}(\omega)\right]^{-2}\to 0$, and $\phi_{opt}(\omega)\simeq -\arctan\left[2/S_{XY}^{out}(\omega)\right]/2 \to 0$.
Since the field quadrature $\delta X^{out}$ at $\phi=0$ is just at the shot-noise limit (see Eq.~(\ref{eq:sxy})), one has that squeezing is achieved only within a narrow interval for the homodyne phase around $\phi_{opt}(\omega)$, of width $ \sim 2 \left |\phi_{opt}(\omega)\right| \sim \arctan\left|2/ S_{XY}^{out}(\omega)\right|$. This extreme phase dependence is a general and well-known property of quantum squeezing, which is due to the Heisenberg principle: the width of the interval of quadrature phases with noise below the shot-noise limit is inversely proportional to the amount of achievable squeezing.

\begin{figure}
\includegraphics[width=1.0\columnwidth]{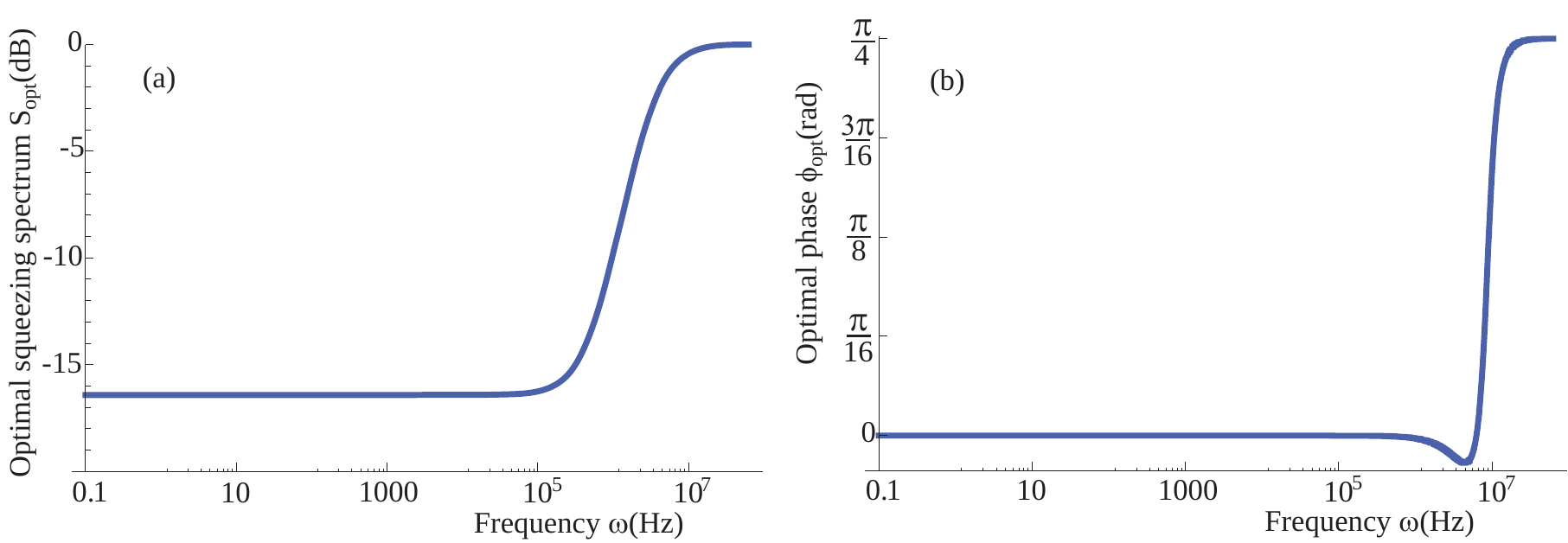}
\caption{Optimal spectrum of squeezing in dB $S_{opt}$ (a), and the corresponding optimal quadrature phase $\phi_{opt}$ (b), versus frequency in the case of a cavity with bandwidth $\kappa = 1$ MHz, length $L=1$ cm, driven by a laser at $1064$ nm and with input power $\mathcal{P}_{in} =10$ mW. The mechanical resonator has $\OmegaM/2\pi= 1$ MHz, mass $\mass=100$ ng, quality factor $Q=10^4$, and temperature $T=4$ K. }
\label{fig:NonClassicalFigure-VitaliTomb}       
\end{figure}

$S_{opt}(\omega)$ and the corresponding optimal phase $\phi_{opt}(\omega)$ at which best squeezing is attained for each $\omega$, are plotted in Fig.~\ref{fig:NonClassicalFigure-VitaliTomb} for a realistic set of parameter values (see figure caption). $S_{opt}(\omega)$ is below the shot-noise limit whenever $S_{XY}^{out}(\omega) \neq 0$ (see Eqs.~(\ref{eq:speopt})-(\ref{eq:sy})), and one gets significant squeezing at low frequencies, well below the mechanical resonance, where the optomechanical cavity becomes fully equivalent to a Kerr medium, as witnessed also by the fact that $\phi_{opt}(\omega)$ is constant in this frequency band. This equivalence is lost close to and above the mechanical resonance, where squeezing vanishes because ${\rm Re}\left\{\chi_M(\omega)\right\} \sim S_{XY}^{out}(\omega) \sim 0$, and the optimal phase shows a large variation.

The present treatment neglects technical limitations: in particular it assumes the ideal situation of a one-sided cavity, where there is no cavity loss because all photons transmitted by the input-output mirror are collected by the output mode. We have also ignored laser phase noise which is typically non-negligible at low frequencies where ponderomotive squeezing is significant. In current experimental schemes both cavity losses and laser phase noise play a relevant role and in fact, ponderomotive squeezing has not been experimentally demonstrated yet. However, the results above show that when these technical limitations will be solved, cavity optomechanical systems may become a valid alternative to traditional sources of squeezing such as parametric amplifiers and Kerr media.

%% file: Part2_Genes.tex
\subsection{EPR correlated beams of light}\label{nonclassical:light:genes}

Optomechanical cavities provide a source not only of squeezed light but also of entangled light, as we will explain in the following. By means of spectral filters, the continuous wave field emerging from the cavity
can be split in many traveling modes thus offering the option
of producing and manipulating a multipartite system
\cite{Genes2008a}. In particular we focus on detecting
the first two motional sidebands at frequencies $\omegaopt\pm\OmegaM$ and show that they posses quantum correlations
of the Einstein-Podolsky-Rosen type \cite{Nonclassical:Einstein1935}.

Using the well-known input-output fields connection
$\ah^{out}(t)=\sqrt{2\kappa}\ah(t)-\ah^{in}(t)$, the output mode can
be split in $N$ independent optical modes by frequency selection
with a proper choice of a causal filter function:
\begin{equation}
\ah^{out}_k(t)=\int_{-\infty}^{t}ds
g_k(t-s)\ah^{out}(s),\;\;\;k=1,\ldots N,\label{filter1},
\end{equation}
where $g_k(s)$ is the causal filter function defining the $k$-th
output mode. The annihilation operators describe $N$ independent
optical modes when
$\left[\ah^{out}_j(t),\ah^{out}_k(t)^{\dagger}\right]=\delta_{jk}$,
which is fulfilled when $\int_{0}^{\infty}ds g_j(s)^*
g_k(s)=\delta_{jk}\label{filter2}$, i.e., the $N$ filter functions
$g_k(t)$ form an orthonormal set of square-integrable functions in
$[0,\infty)$. As an example of a set of functions that qualify as
casual filters we take
\begin{equation}
g_k(t)=\frac{\theta(t)-\theta(t-\tau)}{\sqrt{\tau}}e^{-i\Omega_k t}
, \label{filterex}
\end{equation}
($\theta$ denotes the Heavyside step function) provided that
$\Omega_k$ and $\tau$ satisfy the condition
$\Omega_j-\Omega_k=\frac{2\pi}{\tau}p$ for integer $p$.
Such filtering is seen as a simple frequency integration around
$\Omega_k$ of bandwidth $\sim 1/\tau$ (the inverse of the time
integration window).

For characterization of entanglement one can compute the stationary
$(2N+2)\times (2N+2)$ correlation matrix of the output modes defined
as
\begin{equation}
V^{out}_{ij}(t)=\frac{1}{2}\left\langle u^{out}_i(t)
u^{out}_j(t)+u^{out}_j(t) u^{out}_i(t)\right\rangle,\label{defVout}
\end{equation}
where
\begin{eqnarray}
u^{out}(t)=\left(\hat{q}(t),
\hat{p}(t),\hat{X}_1^{out}(t),\hat{Y}_1^{out}(t),\ldots,\hat{X}_N^{out}(t),\hat{Y}_N^{out}(t)\right)^T
\nonumber
\end{eqnarray}
is the vector formed by the mechanical position and momentum
fluctuations and by the amplitude
($\hat{X}_k^{out}(t)=[\ah^{out}_k(t)+\ah^{out}_k(t)^{\dagger}]/\sqrt{2}$),
and phase
($\hat{Y}_k^{out}(t)=[\ah^{out}_k(t)-\ah^{out}_k(t)^{\dagger}]/i\sqrt{2})$
quadratures of the $N$ output modes.

\begin{figure}[t]
\includegraphics[width=0.99\columnwidth]{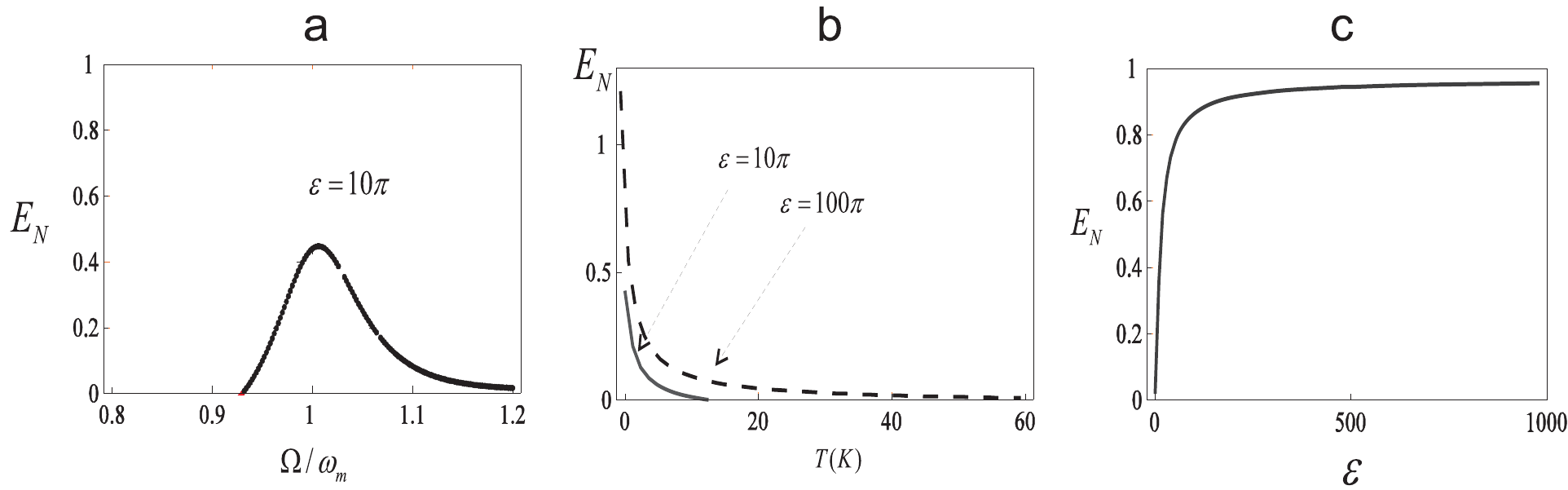}
\caption{a) Logarithmic negativity of Stoke-Antistokes output modes
when $\Omega_1=-\OmegaM $ while $\Omega_2$ is varied around
$\OmegaM$. The inverse bandwidth is kept constant at
$\epsilon=10\pi$. b) Temperature robustness of bipartite
entanglement of output modes at $\pm\OmegaM$ computed for short
($\epsilon=10\pi$, dashed line) and long ($\epsilon=100\pi$, solid
line) detection times. c) The bipartite Stoke-Antistoke entanglement
shows improvement and eventually saturates with increasing the
integration time. Parameters are $\OmegaM/2\pi =10$ MHz, $Q=10^5$, mass
$\mass=50$ ng, cavity of length $\cavlength=1$ mm with finesse
$\finesse=2 \times 10^4$, detuning $\Delta =\OmegaM$, input power $\Pin=30$ mW at $810$ nm, and temperature $T=0.4$ K,
yielding $\gomzero=0.43$ kHz, $\gom=0.41 \OmegaM$, a cavity
bandwidth $\kappa=0.75 \OmegaM$, and a thermal occupation of $\nth \simeq
833$.} \label{nonclassical:genes:fig:1}
\end{figure}

We are now in position to analyze the quantum correlations between
two output modes with the same bandwidth $\tau^{-1}$ and central
frequencies $\Omega_1$ and $\Omega_2$. As a measure for entanglement we apply the
logarithmic negativity $E_{\mathcal{N}}$ to the
covariance matrix of the two optical modes. It is defined as
$E_{\mathcal{N}}=\max [0,-\ln 2\eta ^{-}]$, where $\eta ^{-}\equiv
2^{-1/2}\left[ \Sigma (V)-\left[ \Sigma (V)^{2}-4\det V\right]
^{1/2}\right] ^{1/2}$, with $\Sigma (V)\equiv \det V_{m}+\det
V_{c}-2\det V_{mc}$, and we have used the $2\times 2$ block form of
the covariance matrix
\begin{equation}
V\equiv \left(
\begin{array}{cc}
V_{m} & V_{mc} \\
V_{mc}^{T} & V_{c}%
\end{array}%
\right) .  \label{blocks}
\end{equation}%
Therefore, a Gaussian state is entangled if and only if $\eta
^{-}<1/2$, which is equivalent to Simon's necessary and sufficient
entanglement non-positive partial transpose criterion for Gaussian
states, which can be written as $4\det V<\Sigma -1/4$.

The resulting quantum correlations among the upper and the lower sideband in the continuous wave output field are illustrated in Fig.~\ref{nonclassical:genes:fig:1}. We plot the interesting and not unexpected behavior of
$E_{\mathcal{N}}$ as a function of central detection frequency
$\Omega$ in Fig.\ref{nonclassical:genes:fig:1}a, with the mirror
reservoir temperature in Fig.\ref{nonclassical:genes:fig:1}b and
with the scaled time integration window $\epsilon=\OmegaM\tau$ in
Fig.\ref{nonclassical:genes:fig:1}c. The conclusion of
Fig.\ref{nonclassical:genes:fig:1}a is that indeed scattering off
the mirror can produce good Stokes-Antistokes entanglement which can
be optimized at the cavity output by properly adjusting the
detection window. Moreover, further optimization is possible via an
integration time increase as suggested by
Fig.\ref{nonclassical:genes:fig:1}c. The temperature behavior
plotted in Fig.\ref{nonclassical:genes:fig:1}b shows very good
robustness of the mirror-scattered entangled beams that suggests
this mechanism of producing EPR entangled photons as a possible alternative to
parametric oscillators. 

%% file: Part3_Genes.tex
\subsection{State transfer}
 \label{nonclassical:mechanics:genes}

For a massive macroscopic mechanical resonator, just as in the case
of a light field, the signature of quantum can be indicated in a
first step by the ability of engineering of a squeezed state. Such a
state would also be useful in ultrahigh precision measurements or
detection of gravitational waves and has been experimentally proven
in only one instance for a nonlinear Duffing resonator
\cite{Nonclassical:Almog2007}. Numerous proposals exist and can be
categorized as i) direct: modulated drive in optomechanical settings
with or without feedback loop
\cite{Nonclassical:Clerk2008,Nonclassical:Vitali2002,Nonclassical:Woolley2008,Nonclassical:Tian2008},
and ii) indirect: mapping a squeezed state of light or atoms onto
the resonator, coupling to a cavity with atomic medium within
\cite{Nonclassical:Ian2008}, coupling to a Cooper pair box
\cite{Nonclassical:Rabl2004} or a superconducting quantum
interference loop
\cite{Nonclassical:Zhou2006,Nonclassical:Zhang2009}. In the
following we take the example of state transfer in a pure
optomechanical setup where laser cooling of a mirror/membrane via a
strong laser is accompanied by squeezing transfer from a squeezed
vacuum second input light field \cite{Nonclassical:Jahne2009}. While
the concept is straightforward it is of interest to answer a few
practical questions such as: i) what is the resonance condition for
optimal squeezing transfer and how does a frequency mismatch affect
the squeezing transfer efficiency, ii) what is the optimal transfer,
iii) how large should the cavity finesse should be for optimal
transfer etc.

\begin{figure}[t]
\includegraphics[width=0.99\columnwidth]{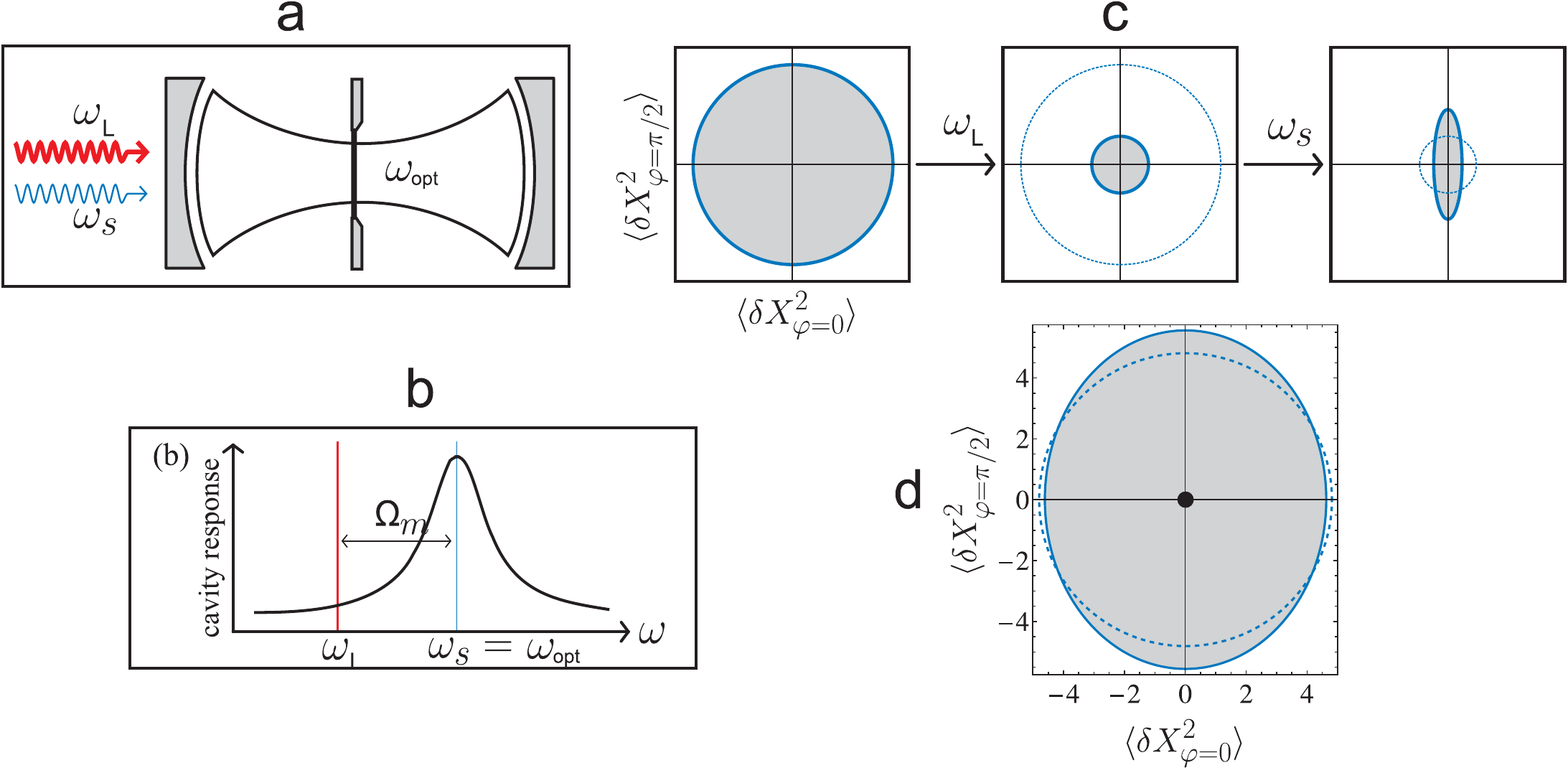}
\caption{a) Logarithmic negativity of Stoke-Antistokes output modes
when $\Omega_1=-\OmegaM $ while $\Omega_2$ is varied around
$\OmegaM$. The inverse bandwidth is kept constant at
$\epsilon=10\pi$. b) Temperature robustness of bipartite
entanglement of output modes at $\pm\OmegaM$ computed for short
($\epsilon=10\pi$, dashed line) and long ($\epsilon=100\pi$, solid
line) detection times. c) The bipartite Stoke-Antistoke entanglement
shows improvement and eventually saturates with increasing the
integration time.} \label{nonclassical:genes:fig:2}
\end{figure}

To this purpose we assume the typical membrane in the middle
optomechanical system depicted in Fig.
\ref{nonclassical:genes:fig:2}a, where the laser Antistokes
sideband, the cavity and the squeezing central frequency are
resonant (shown in \ref{nonclassical:genes:fig:2}b). The input
squeezed light operators have the following correlations
\begin{eqnarray}
\langle\hat{c}_{\rm in}(t+\tau)\hat{c}_{\rm in}(t)\rangle=\frac{M}{2}\frac{b_x b_y}{b_x^2+b_y^2}\left(b_y e^{-b_x|\tau|}+b_x e^{-b_y |\tau|}\right)\\
\langle \hat{c}^\dag_{\rm in}(t+\tau)\hat{c}_{\rm
in}(t)\rangle=\frac{N}{2}\frac{b_x b_y}{b_y^2-b_x^2}\left(b_y
e^{-b_x|\tau|}-b_x e^{-b_y |\tau|}\right).
\end{eqnarray}
The noise operators are written in a frame rotating at $\omega_{\rm
s}$ and satisfy the canonical commutation relation $[ \hat{c}_{\rm
in}(t),\hat{c}^\dag_{\rm in}(t')]=\delta(t-t')$. Parameters $N$ and
$M$ determine the degree of squeezing, while $b_x$ and $b_y$ define
the squeezing bandwidth. For pure squeezing there are only two
independent parameters, as in this case $|M|^2=N(N+1)$ and
$b_y=b_x\sqrt{2\left(N+|M|\right)+1}$.

Following the standard linearized quantum Langevin equations
approach for optomechanics, we first identify two conditions for
optimal squeezing: i) $\Delta =\OmegaM$, meaning that we require
continuous laser cooling in the resolved sideband regime and ii)
$\Delta_s =-\OmegaM$ so that the squeezing spectrum is centered
around the cavity frequency. Then we look at the variances of the
generalized quadrature operator
\begin{eqnarray}
\delta \hat{X}_{\varphi}(t)&=&\frac{1}{\sqrt{2}}\left(e^{i\varphi}
\hat{b}(t)+e^{-i\varphi}
\hat{b}^\dag(t)\right)\label{Eq:GenQuadrature},
\end{eqnarray}
which for $\varphi=0$ is the usual position operator $\hat{q}(t)$
and for $\varphi=-\pi/2$ is the momentum operator $\hat{p}(t)$, both
taking in a rotating frame at frequency $\OmegaM$. In the limit of
squeezed white noise the quadrature correlations take a simple form
\begin{eqnarray}
\langle \delta \hat{X}_{\varphi}(t) \delta
\hat{X}_{\varphi}(t)\rangle= \left(N+\frac{1}{2}-{\rm Re}\left\{M
e^{2i\varphi}\right\}\right)
+\frac{\GammaM}{\GammaEff}\left(\nth+\frac{1}{2}\right)\label{Eq:XXCorrRS}.
\end{eqnarray}
The first term in the right hand side comes from the squeezing
properties of the squeezed input vacuum while the second term is the
residual occupancy after laser cooling. In view of this equation a
successful squeezed mechanical state preparation automatically
requires close to ground state cooling. One can follow this in
Fig.\ref{nonclassical:genes:fig:2}c where cooling close to ground
state of an initially thermal mechanical state is performed by the
cooling laser, and subsequently squeezing of a quadrature is
achieved via the squeezed vacuum. In
Fig.\ref{nonclassical:genes:fig:2}d we suggest that although ground
state cooling is not easy to achieve, one can at least in principle
(for example by complete tomography of the mechanical state) see a
quadrature squashing effect.

To answer a practical question, when the squeezing is not white,
fulfilling the resonance condition $\Delta_s =-\OmegaM$ is
important. The deviations allowed for the frequency mismatch are
smaller than the width of the cooling sideband, i.e. $\GammaEff$. A
second question is the effect of the finite width of squeezing. In
general there is an optimal squeezing bandwidth for which the
transfer from light to membrane is maximized, but in the resolved
sideband limit where $\OmegaM\gg\kappa$ the finite bandwidth result
does not differ much from the infinite bandwidth limit result. For a
large bandwidth which fully covers the motional sidebands, $b_x \gg
\OmegaM$, the membrane sees only white squeezed input noise, whereas
for smaller bandwidth, the crucial question is whether the squeezed
input will touch the heating sideband or not. For a high-finesse
cavity, the width is not a big issue, since the heating sideband is
anyway weak, whereas for a bad cavity the squeezing transfer is much
improved for an optimal, finite bandwidth where the strong heating
sideband is avoided.

%% file: Part4_Milburn.tex
\subsection{Continuous measurements of mechanical oscillators. }

Another method for creating nonclassical states exploits the possibility to conditionally prepare states of the mechanical oscillator via measuring the output field of the optomechanical system.
The coupling of a mechanical resonator to an optical cavity field enables an indirect continuous monitoring of the mechanical motion by a direct phase dependent measurement of the field leaving the cavity. The standard radiation pressure coupling is linear in the displacement of the mechanical resonator and thus enables a continuous measurement of displacement. If we wish to monitor the energy (or phonon number) of the mechanical resonator, however, we need to find an interaction Hamiltonian that is quadratic in the displacement.  Such interactions can occur in a number of ways \cite{Sankey2010,Romero-Isert2011}. We begin with the case of displacement measurements.

The standard linearised opto-mechanical coupling Hamiltonian
\begin{equation}
H=\hbar\Delta \ah^\dagger \ah+\hbar\OmegaM \bh^\dagger \bh-\hbar \gom(\ah+\ah^\dagger)(\bh +\bh^\dagger)
\end{equation}
As the interaction part of this Hamiltonian commutes with the (dimensionless) mechanical displacement operator, $\hat{q}=\frac{1}{\sqrt{2}}(\bh+\bh^\dagger)$,  in principle this model can be configured as a measurement of the displacement.  However as $\hat{q}$ does not commute with the free mechanical Hamiltonian, this is not a strict QND measurement \cite{Caves1980}. Nonetheless, for a rapidly damped cavity, we can effect can approximate QND readout of the mechanical displacement provided the coupling constant $\gom$ can be turned on and off sufficiently fast. This can be achieved by using a pulsed coherent driving field on the cavity \cite{Vanner2011}.

If we include the damping of both the cavity and the mechanics, we obtain the quantum stochastic differential equations,
\begin{eqnarray}
\frac{d\ah}{dt} & = & -i\Delta \ah-\frac{\kappa}{2}\ah+i\gom\hat{X}+\sqrt{\kappa}\ah_{in}\label{a_qsde}\\
\frac{d\bh}{dt} & = & -i\OmegaM \bh-\frac{\GammaM}{2}\bh+i\gom(\ah+\ah^\dagger)+\sqrt{\GammaM}\bh_{in}\label{b_qsde}
\end{eqnarray}
where we assume that the input noise to the cavity is vacuum, so that the only non zero correlation function for the cavity noise is $\langle \ah_{in}(t)\ah^\dagger_{in}(t')\rangle=\delta(t-t')$ but that the input noise to the mechanical resonator is thermal, $\langle \bh_{in}(t)\bh_{in}^\dagger(t')\rangle=(\nth+1) \delta(t-t')$. We expect that a good measurement will occur when the cavity field is rapidly damped so that it is slaved to the mechanical degree of freedom. We can then adiabatically eliminate the cavity degree of freedom by setting to zero the right hand side of Eq.({\ref{a_qsde}) and formally solving for the operator $\ah$,
\begin{equation}
\ah= \frac{ige^{-i\phi}}{\sqrt{\Delta^2+\kappa^2/4}}\hat{q}+\frac{\sqrt{\kappa}}{i\Delta+\kappa/2}\ah_{in}
\end{equation}
where $\tan\phi=2\Delta/\kappa$. The actual output field from the cavity is related to the field inside by the input/output relation, $\ah_{out}=\sqrt{\kappa}\ah-\ah_{in}$, so that
\begin{equation}
\ah_{out}(t)=\frac{ig\sqrt{\kappa}e^{-i\phi}}{\sqrt{\Delta^2+\kappa^2/4}}\ \hat{q}(t)+e^{-2i\phi}\ah_{in}(t)
\end{equation}
Clearly this indicates that we need to measure a particular quadrature of the output field (for example by homodyne or heterodyne detection) and that the added noise is vacuum noise. The optimal transfer is obtained on resonance.  A fast, impulsive readout of the mechanical resonator's displacement may be made by injecting a coherent pulse into the cavity and subjecting the output pulse to a homodyne measurement \cite{Vanner2011}.

If we wish to measure the energy of a mechanical resonator we must find an interaction hamiltonian that is at least quadratic in the mechanical amplitude. A number of schemes have been proposed, including trapped atoms in a standing wave \cite{Domokos2003} and  a nanomechanical resonator coupled to a single superconducting junction dipole in the dispersive regime \cite{Woolley2010}.  In opto-mechanics  a dielectric membrane placed at the antinode of a cavity standing wave shifts the cavity frequency proportional to the square of the mechanical displacement of the membrane from equilibrium \cite{Sankey2010}. A similar interaction arises for an optically levitated particle in a standing wave \cite{Romero-Isert2011}.

The interaction Hamiltonian in this case takes the form
\begin{equation}
H=\hbar\omega_{opt}\ah^\dagger \ah+\hbar\OmegaM \bh^\dagger \bh+\hbar(\epsilon_c^* \ah e^{i\omega_L t}+\epsilon_c \ah^\dagger e^{-i\omega_l t})
+\frac{\hbar}{2} G_2 \ah^\dagger \ah(\bh+\bh^\dagger)^2,
\label{ham2}
\end{equation}
where
\begin{equation}
G_2=\frac{\hbar}{2\nu m}\left . \frac{\partial^2\omega_c(x)}{\partial x^2}\right |_{x=x_0}.
\end{equation}
and where we have included a coherent driving field with amplitude $\epsilon$. As usual we expand the interaction around the steady state cavity field.  After the rotating wave approximation, the effective Hamiltonian in the interaction picture may then be written as
\begin{equation}
H_I=\frac{\hbar}{2}\chi(\bar{\ah}+\bar{\ah}^\dagger)\bh^\dagger \bh,
\label{ham-int}
\end{equation}
where $\chi=2G_2\alpha_0$.

The interaction in Eq.(\ref{ham-int}) describes a displacement of the cavity field proportional to the number of vibrational quanta in the mechanical resonator. The average steady-state displacement is given by $\chi\bar{n}/\kappa$, where $\bar{n}$ is the mean phonon number operator for the mechanical oscillator, $\kappa$ is the cavity line-width.
 If we continuously monitor the output field amplitude from the cavity via homodyne detection this scheme can in principle enable a continuous monitoring of the mechanical vibrational energy, and phonon number jumps \cite{Gangat2011}.

Under continuous homodyne measurement of this quadrature, the system is governed by the following stochastic master equation (SME):
\begin{equation}
\label{SME}
d\rho=-\frac{i}{\hbar}[H_I,\rho]dt + \gamma(\bar{n}_{th}+1)\mathcal{D}[\bh]\rho dt + \gamma\bar{n}_{th}\mathcal{D}[\bh^\dag]\rho dt + \kappa\mathcal{D}[\ah]\rho dt + \sqrt{\kappa}dW\mathcal{H}[ae^{-i\frac{\pi}{2}}]\rho,
\end{equation}
where ${\cal D}[c]\rho=c\rho c^\dagger -c^\dagger c\rho/2-\rho c^\dagger c/2$ and $\mathcal{H}[c]\rho=c\rho + \rho c^\dag -$ Tr$(c\rho + \rho c^\dag)$ is the measurement super-operator, $\gamma$ and $\kappa$ are the respective mechanical and cavity damping rates.

In figure \ref{fig_SRE_18-20} we show a numerical integration of the stochastic master equation with $\kappa=10^4\gamma\bar{N}$ for three cases: $\chi^2/\kappa=\gamma\bar{N}$, $\chi^2/\kappa=10\gamma\bar{n}_{th}$, and $\chi^2/\kappa=10^2\gamma\bar{n}_{th}$.  We start with the mechanics in the ground state, and the bath temperature is set at $\bar{n}_{th}=0.5$.  The first case, Fig. (\ref{fig_SRE_18-20}a), does not satisfy the fast-measurement condition and therefore does not resolve quantum jumps in the phonon number.  The second case, Fig. (\ref{fig_SRE_18-20}b), is on the border of the fast-measurement regime for $n\sim1$ and shows some jump-like behaviour in the phonon number.  The third case, Fig. (\ref{fig_SRE_18-20}c), strongly satisfies the fast-measurement condition for low phonon numbers and shows well-resolved quantum jumps in spite of being deeply within the weak coupling regime with $\chi/\kappa=10^{-1}$.

\begin{figure} [htp]
\begin{center}
\includegraphics[scale=0.6]{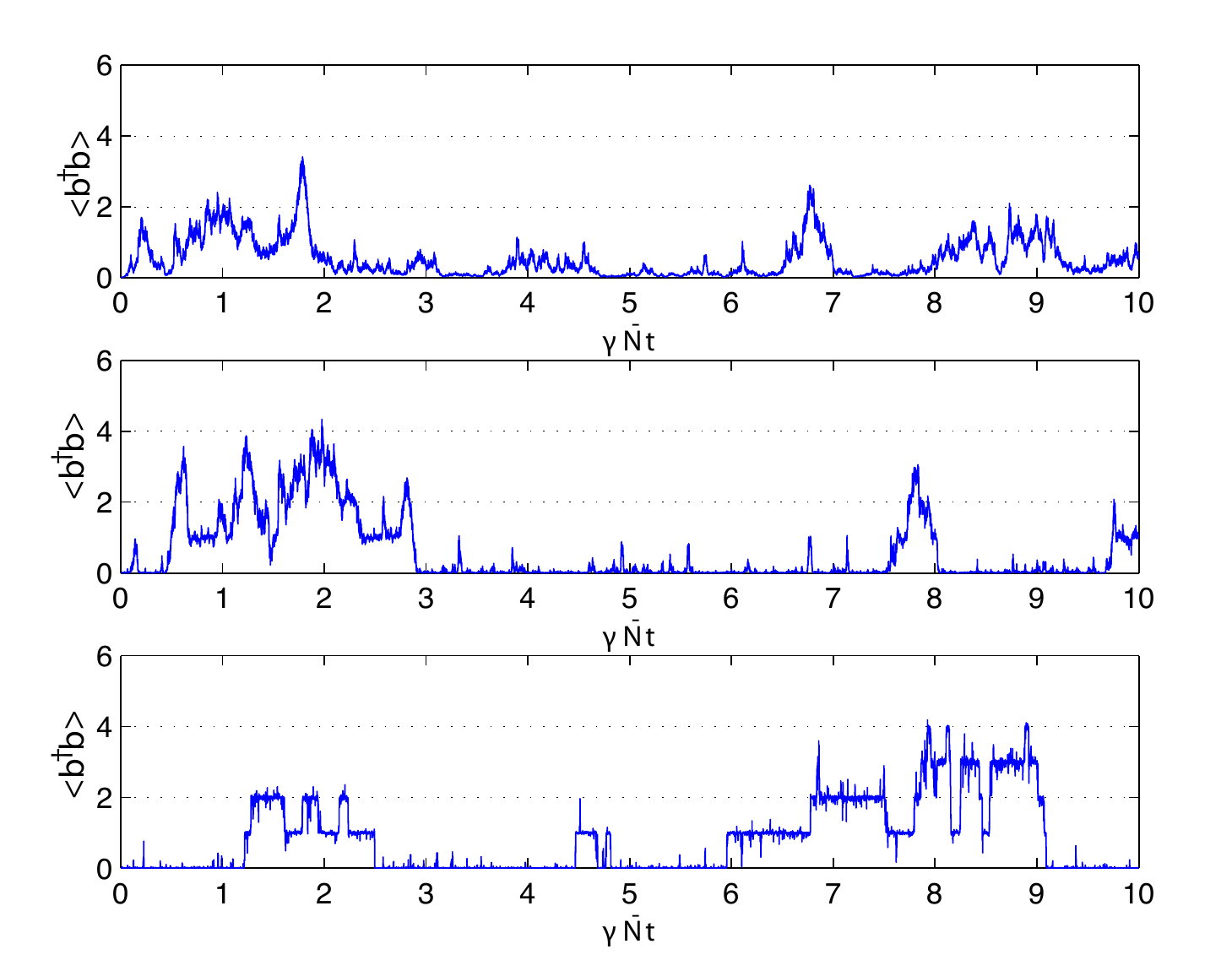}
\end{center}
\caption{The evolution of the conditional average phonon number with parameters $\kappa=10^4\gamma\bar{n}_{th}$ and: (a) $\chi^2/\kappa=\gamma\bar{n}_{th}$, (b) $\chi^2/\kappa=10\gamma\bar{N}$, (c) $\chi^2/\kappa=10^2\gamma\bar{n}_{th}$.  Jump-like behaviour occurs only when $\chi^2/\kappa\gg\gamma[\bar{n}_{th}(n+1)+(\bar{n}_{th}+1)n]$, where $n$ is the phonon number.}
\label{fig_SRE_18-20}
\end{figure}

For jump-like behaviour to arise in the weak coupling limit, the adiabatic condition is not sufficient.  In this regime, analysis shows that the rate of information acquisition about the phonon number is proportional to $\chi^2/\kappa$.  As in the strong coupling case, this measurement rate must dominate the thermalisation rate in order for quantum jumps to arise.  Thus, in addition to being in the adiabatic limit, the weak coupling regime requires $\chi^2/\kappa\gg\gamma\bar{N}$.

The state of the mechanical system conditioned on the measurement of light will be in a Non-Gaussian state. This is due to the non-linear interaction introduced in \eqref{ham2}. Another way of achieving a Non-Gaussian state exploits the nonlinearity provided in photon counting, as will be detailed in the next section.

%% file: Part5_Bouwmeester.tex
%

In $1935$ Schr\"{o}dinger pointed out that according to quantum
mechanics even macroscopic systems can be in superposition states
\cite{Schroedinger1935}. The interference effects, characteristic of
quantum mechanics, are expected to be hard to detect due to
environment induced decoherence \cite{decoherence}. Nevertheless
there have been several proposals on how to create and observe
macroscopic superpositions in various physical systems. See references
\cite{Ruostekoski1998, Bose1999, Armour2002} for some of the first proposals. There have also been experiments
on superposition states in superconducting and piezoelectric devices \cite{VanderWaletal2000, OConnell2010}
and on interference with fullerene \cite{Arndt1999} and other large molecules. One
long-term motivation for this kind of experiment is the question
of whether unconventional decoherence processes such as
gravitationally induced decoherence or spontaneous wave-function
collapse \cite{Ghirardi1986, Ghirardi1990, Fivel1997, Percival1994, Penrose2000} take place.

In this section a scheme is analyzed that is close in spirit to
Schr\"{o}dinger's original discussion. A small quantum system (a
photon) is coupled to a large system (a mirror) such that they
become entangled \cite{Marshall2003}. The system consists of a
Michelson interferometer in which one arm has a tiny moveable
mirror. The radiation pressure of a single photon is used to
displace the tiny mirror. The initial superposition of the photon
being in either arm causes the system to evolve into a
superposition of states corresponding to two distinct spatial
locations of the mirror. A high-finesse
cavity is used to enhance the interaction between the single
photon and the mirror. The interference of the photon upon exiting the interferometer
allows one to study the creation of coherent superposition states
periodic with the motion of the mirror.

\begin{figure}
\begin{center}
\includegraphics[width=0.5\columnwidth]{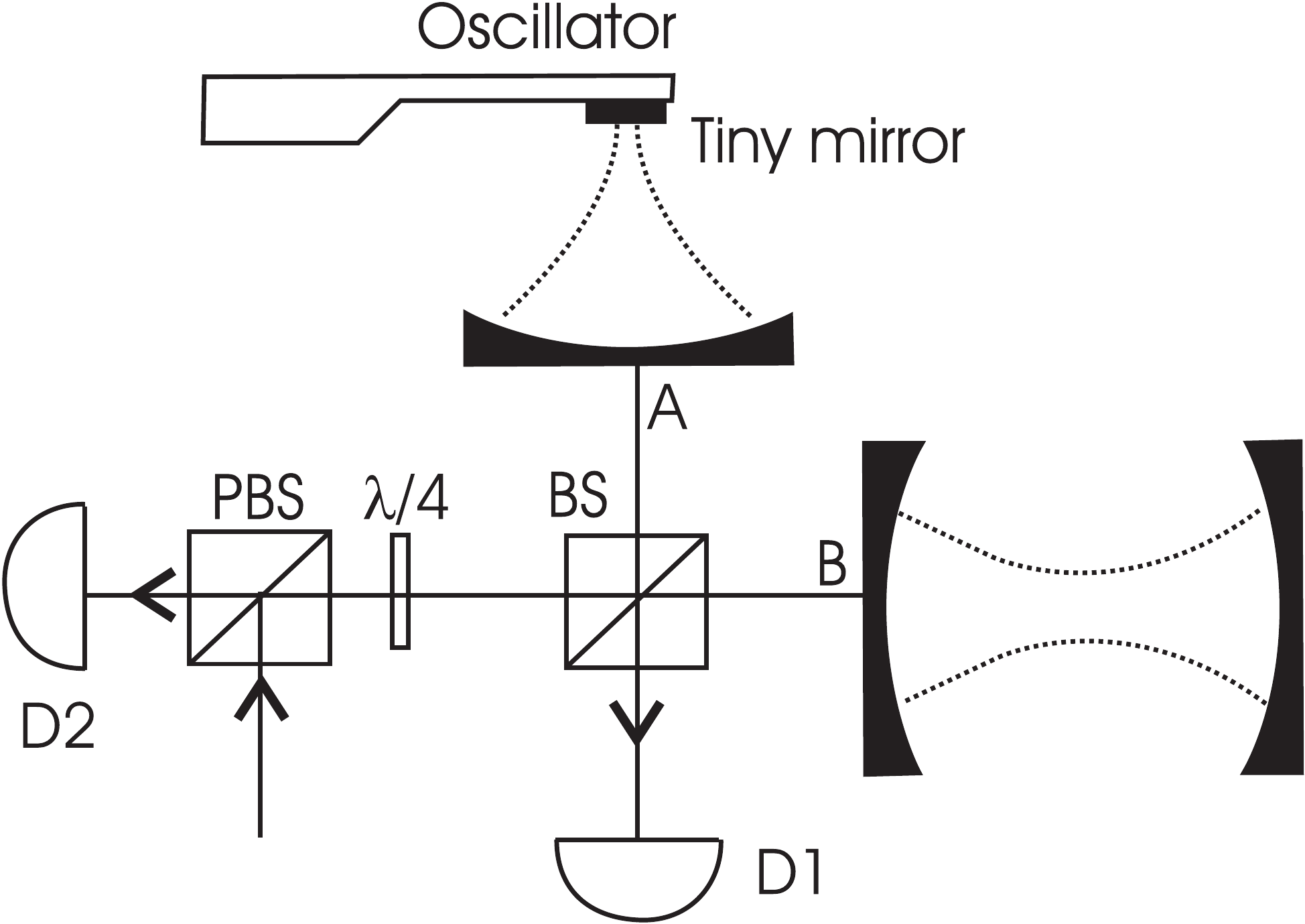}
\caption{Scheme: a Michelson interferometer for a
single photon with in each arm a high-finesse cavity.
The cavity in arm A has a very small end mirror mounted on a
micro-mechanical oscillator. The single photon enters the interferometer via a polarizing beam splitter (PBS) followed by a \(\lambda/4\) wave plate. This is an optical trick to allow detection of the photon leaking out of the interferometer at a later time on detector D1 or D2.
If the input photon is considered to be in arm A, the motion of the small mirror is
effected by its radiation pressure. If the input photon is considered to be in arm B, the motion of the mirror is undisturbed. The interferometer, based on the 50\%/50\% beamsplitter (BS), leads to the entanglement between the photon being in arm A or in arm B and the state of the mirror.}
\label{setup}
\end{center}
\end{figure}

Consider the setup shown in Fig.~\ref{setup}, consisting of a Michelson
interferometer that has a cavity in each arm. In the cavity in arm
A one of the mirrors is very small and attached to a
micromechanical oscillator. While the photon is inside the cavity,
it exerts a radiation pressure force on the small mirror. We will
be interested in the regime where the period of the mirror's
motion is much longer than the roundtrip time of the photon inside
the cavity, and where the amplitude of the mirror's motion is very
small compared to the cavity length. Under these conditions, the
system can be described by the standard optomechanical Hamiltonian \cite{Law1993, Law1994}
\begin{eqnarray}
H=\omegaopt \ah^{\dagger} \ah + \OmegaM \bh^{\dagger} \bh  - \gomzero
\ah^{\dagger} \ah (\bh+\bh^{\dagger}).
\end{eqnarray}
 To start
with, let us suppose that initially the photon is in a
superposition of being in either arm $A$ or $B$, and the mirror is
in a coherent state $|\beta \rangle = e^{-|\beta|^{2}/2} \sum
\limits_{n=0}^{\infty} \frac{\beta^{n}}{\sqrt{n!}}|n\rangle$, where
$|n\rangle$ are the eigenstates of the harmonic oscillator. Then the
initial state is
\begin{equation} |\psi(0)\rangle=\frac{1}{\sqrt{2}}(|0\rangle_{A} |1\rangle_{B}
+ |1\rangle_{A} |0\rangle_{B})|\beta\rangle. \end{equation}
After a time $t$ the state of the
system will be given by \cite{Mancini1997, Bose1997}
\begin{align}\label{state}
|\psi(t)\rangle& =\frac{1}{\sqrt{2}}e^{-i\omegaopt t} \{ |0\rangle_A |1\rangle_B |\beta
e^{-i \OmegaM t}\rangle\nonumber\\
&\quad+ e^{i\eta^2(\OmegaM t-\sin
\OmegaM t)}|1\rangle_A |0\rangle_B  |\beta e^{-i \OmegaM t}+
\eta(1-e^{-i \OmegaM t})\rangle \},
\end{align}
where $\eta=\gomzero/\OmegaM$. In the second term on the right hand side the
motion of the mirror is altered by the radiation pressure of the
photon in cavity $A$. The parameter $\eta$ quantifies the
displacement of the mirror in units of the size of the coherent
state wavepacket. In the presence of the photon the mirror
oscillates around a new equilibrium position determined by the
driving force.

The maximum possible interference visibility for the photon is
given by twice the modulus of the off-diagonal element of the
photon's reduced density matrix. By tracing over the mirror one
finds from Eq. (\ref{state}) that the off-diagonal element has the
form \begin{eqnarray}
\frac{1}{2} e^{-\eta^2(1-\cos \OmegaM t)}
e^{i\eta^2(\OmegaM t-\sin \OmegaM t)+i \eta \mbox{\small
Im}[\beta (1-e^{i\OmegaM t})]}
\label{coherence} \end{eqnarray}
where Im denotes the imaginary part. The first factor is the modulus,
obtaining its minimum value after half a period at
$t=\pi/\OmegaM$, when the mirror is at its maximum displacement.
The second factor gives the phase, which is identical to that
obtained classically due to the varying length of the cavity.

For general $t$ the phase in Eq. (\ref{coherence}) depends on
$\beta$, i.e. the initial condition of the mirror. However, the
effect of the initial condition averages out after every full
period.

In the absence of decoherence, after a full period,
$t=2\pi/\OmegaM$, the system is in the state
$\frac{1}{\sqrt{2}}(|0\rangle_A |1\rangle_B + e^{i\eta^2 2 \pi} |1\rangle_A
|0\rangle_B) |\beta\rangle$, such that the mirror is again completely
disentangled from the photon. Full interference can be observed if
the photon is detected at that moment. If the environment of the
mirror ``remembers'' that the mirror has moved, then, even after a
full period, the photon will still be entangled with the mirror's
environment, and thus the interference for the photon will be
reduced. Therefore the setup can be used to measure the
decoherence of the mirror.

\begin{figure}
\begin{center}
\includegraphics[width= 0.465 \columnwidth]{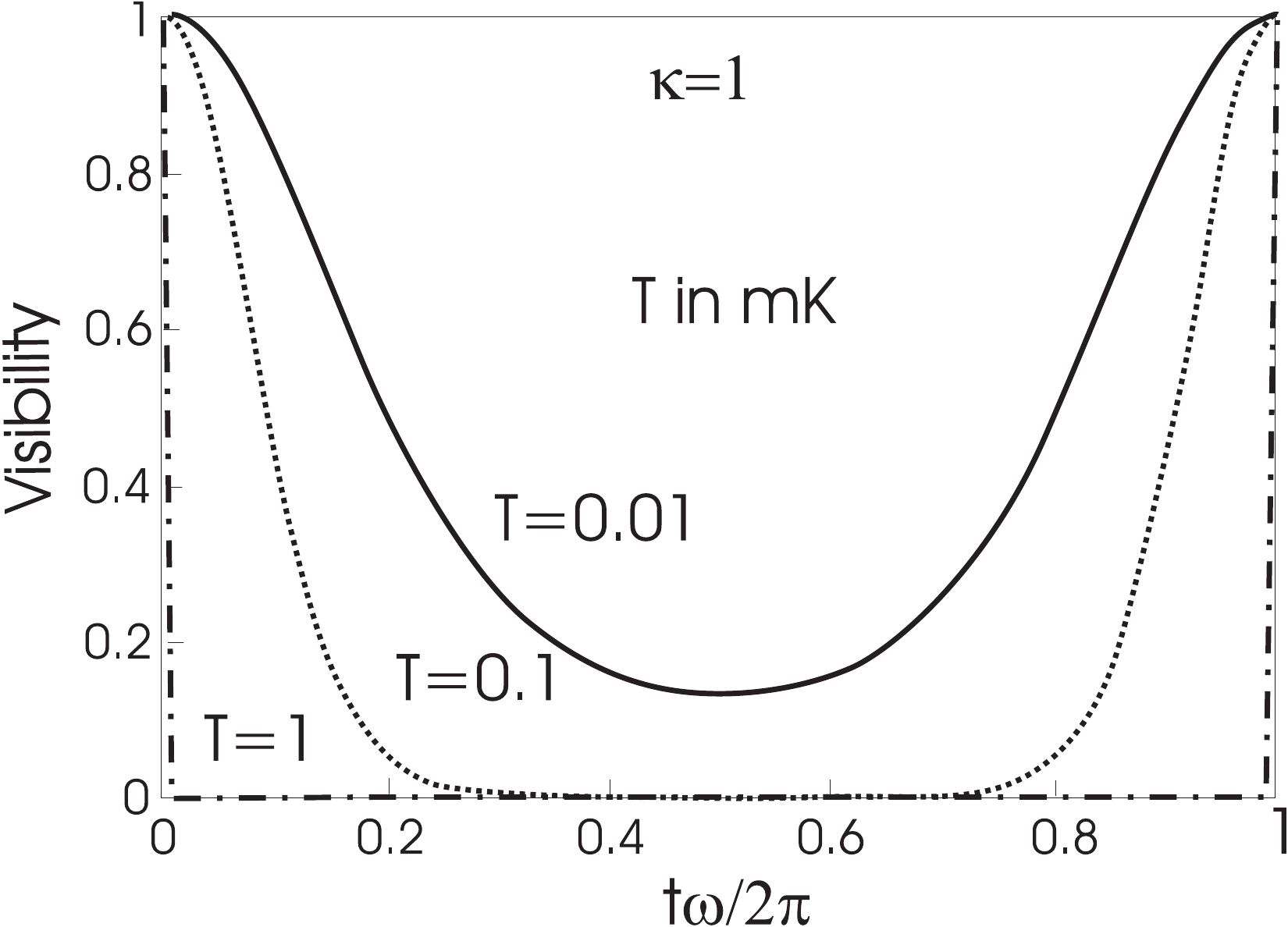}
\caption{Time evolution of the interference visibility of the
photon over one period of the mirror's motion for $\eta=1$ and
temperatures $T$ of 1 mK, 100 $\mu$K and 10 $\mu$K. The visibility
decays very fast after $t=0$, but in the absence of decoherence
there is a sharp revival of the visibility after a full period (2$\pi$).
The width of each peak scales like $1/\sqrt{T}$. } \label{visibility}
\end{center}
\end{figure}

In practice the mirror attached to a mechanical-resonator will be in a thermal state, which
can be written as a mixture of coherent states $|\beta\rangle$ with a
Gaussian probability distribution $(1/\pi
\nth)e^{-|\beta|^2/\nth}$, where $\nth$ is the mean
thermal number of excitations, $\nth=1/(e^{\hbar
\OmegaM/kT}-1)$. If one wants to determine the expected
interference visibility of the photon at a time $t$ for an initial
mirror state which is thermal, one therefore has to average the
off-diagonal element Eq. (\ref{coherence}) over $\beta$ with this
distribution. The result is \begin{eqnarray} \frac{1}{2}
e^{-\eta^2(2\nth+1)(1-\cos \OmegaM t)} e^{i\eta^2(\OmegaM
t-\sin \OmegaM t)}. \label{thermal} \end{eqnarray} As a consequence of the
averaging of the $\beta$-dependent phase in Eq.
({\ref{coherence}}), the off-diagonal element now decays on a
timescale $1/(\eta \OmegaM \sqrt{\nth})$ after $t=0$, i.e.
very fast for the realistic case of large $\nth$. However,
remarkably it still exhibits a revival \cite{Bose1999} at
$t=2\pi/\OmegaM$, when photon and mirror become disentangled and
the phase is independent of $\beta$, such that the phase averaging
does not reduce the visibility. Figure~\ref{visibility} shows the time evolution of the visibility of the
photon over one period of the mirror's motion for $\eta=1$ and
temperatures $T$ of 1 mK, 100 $\mu$K and 10 $\mu$K.

The magnitude of the revival is
reduced by any decoherence of the mirror. Furthermore the revival will also be reduced due to nonlinear terms in the mechanical oscillator. However since we will only consider extremely small displacements around the equilibrium position we assume that nonlinear effects can be ignored.

The revival demonstrates the coherence of the superposition state
that exists at intermediate times. For $\eta^2 \gtrsim 1$ the
state of the system is a superposition of two distinct positions
of the mirror. More precisely, for a thermal mirror state, the
state of the system is a mixture of such superpositions. However,
this affects neither the fundamentally non-classical character of
the state nor, as we have seen, the existence of the revival after
a full period. We now discuss the experimental requirements for
achieving such a superposition state and observing its recoherence
at $t=2 \pi/\OmegaM$.

Firstly, it is required that $\eta^2 \gtrsim 1$, which physically means
that the momentum kick imparted by the photon to the mirror has to
be larger than the initial quantum momentum uncertainty of the
mirror. Let $N$ denote the number of roundtrips of the photon in
the cavity during one period of the mirror's motion, such that
$2NL/c=2\pi/\OmegaM$. This allows us to rewrite the condition
$\eta^2 \gtrsim 1$ as \begin{eqnarray} \frac{2 \hbar N^3 L}{\pi c \mass \lambda^2
} \gtrsim 1, \label{required} \end{eqnarray} where $\lambda$ is the
wavelength of the light. The factors entering Eq. (\ref{required})
are not all independent. The achievable $N$, which is determined
by the quality of the mirrors, and the minimum possible mirror
size depend strongly on $\lambda$. The mirror's lateral dimensions
should be an order of magnitude larger than $\lambda$ to limit
diffraction and avoid geometrical losses. The minimum possible
thickness of the mirror generally depends on the wavelength as
well in order to achieve sufficiently low transmission.

Eq. (\ref{required}) allows one to compare the viability of
different wavelength ranges. While the highest values for $N$ are
achievable for microwaves (up to $10^{10}$), this is counteracted
by their long wavelengths (of order cm). On the other hand there
are no good mirrors for highly energetic photons. The optical
regime seems optimal. In the following estimates we will consider a $\lambda$
around 630 nm.

The cavity mode needs to have a very narrow focus on the tiny
mirror, which requires the other cavity end mirror to be large due
to beam divergence. The maximum cavity length is therefore limited
by the difficulty of making large high quality curved mirrors. In fact from simulations it follows that indeed the surface quality of the large curved mirror is likely to be the most challenging component of the setup \cite{Kleckner2006, Kleckner2010}. Here we consider
a cavity length of 5 cm, and a small mirror size of 10 x 10 x 10
microns, leading to a mass of order $5 \times 10^{-12}$ kg.

One possible path for the fabrication of such a small mirror on a
good mechanical oscillator is to coat a silicon cantilever with
alternating layers of SiO$_{2}$ and a metal oxide such as
Ta$_2$O$_5$ by sputtering deposition. The best current mirrors are
made in this way. Recently such mirrors have also been produce on Silicon Nitride cross resonators which have excellent mechanical properties \cite{Kleckner2011}.

For the above dimensions the condition (\ref{required}) is
satisfied for $N=5.6 \times 10^6$. Therefore a photon loss
per reflection not larger than $3 \times 10^{-7}$ is needed, which is about
a factor of 4 below the best reported values for such mirrors
\cite{Rempe1992}, and for a transmission of $10^{-7}$, which is
consistent with the quoted mirror thickness \cite{Hood2001}. For these
values, about 1\% of the photons are still left in the cavity
after a full period of the mirror. Coupling into a high-finesse
cavity with a tight focus will require carefully designed
incoupling optics. For the above values of $N$ and $L$ one obtains
a frequency $\OmegaM=2 \pi \times 500$ Hz. This leads to a
quantum uncertainty of order $10^{-13}$ m, which for $\eta^2
\sim 1$ corresponds to the displacement in the superposition.

Secondly, the requirement of observing the revival after a full
period of the mirror's motion puts a bound on the acceptable
environmental decoherence. To estimate the expected decoherence we
model the mirror's environment by an (Ohmic) bath of harmonic
oscillators. Applying the analysis of \cite{Strunz2002} one then finds
that off-diagonal elements between different mirror positions
decay with a factor
\begin{eqnarray} \exp{\left[-\frac{\GammaM k T \mass (\Delta
x)^2}{\hbar^2} \left( t+\frac{\sin \OmegaM t \cos \OmegaM
t}{\OmegaM} \right) \right]},\label{decoherence}
\end{eqnarray}
where $\GammaM$ is the rate of energy dissipation for the mechanical
oscillator, $T$ is the temperature (which is constituted mainly by
the internal degrees of freedom of the mirror cantilever) and
$\Delta x$ is the separation of two coherent states that are
originally in a superposition. Note that our experiment is not in
the long-time regime where decoherence is characterized simply by
a rate. However, the oscillatory term in the exponent of Eq.
(\ref{decoherence}) does not affect the order of magnitude and
happens to be zero after a full period. Assuming that the
experiment achieves $\eta^2 \gtrsim 1$, i.e. a separation by the
size of a coherent state wavepacket, $\Delta
x=\sqrt{\frac{\hbar}{\mass\OmegaM}}$, the condition that the exponent
in Eq. (\ref{decoherence}) should be at most of order 1 after a
full period can be cast in the form \begin{eqnarray} Q \gtrsim \frac{kT}{\hbar
\OmegaM}=\nth, \end{eqnarray} where $Q=\OmegaM/\GammaM$ is the quality
factor of the mechanical oscillator. Bearing in mind that quality
factors of the order of $10^5-10^6$ have been achieved for
silicon cantilevers of approximately the right dimensions and
frequency, this implies that the temperature has to be
approximately 3-30 mK. It will be beneficial to perform experiments at even lower temperatures
to reduce the measurement time, as we will explain below.

Thirdly, the stability requirements for the experiment are very
strict. The phase of the interferometer has to be stable over the
whole measurement time. This means that in particular the distance
between the large cavity end mirror and the equilibrium position
of the small mirror has to be stable to of order
$\lambda/20 N=0.6 \times 10^{-14}$m.

The required measurement time can be determined in the following
way. A single run of the experiment starts by sending a weak pulse
into the interferometer, such that on average 0.1 photons go into
either cavity. This probabilistically prepares a single-photon
state as required to a good approximation. The two-photon
contribution has to be kept low because it causes noise in the
interferometer. From Eq. (\ref{thermal}) the width of the revival
peak is $2/\eta \OmegaM \sqrt{\nth}$. This implies that only
a fraction $\sim 1/\pi\sqrt{\nth}$ of the remaining photons
will leak out in the time interval of the revival. It is therefore
important to work at the lowest possible temperature. Temperatures below 100 $\mu$K can be achieved with a nuclear
demagnetization cryostat.

Together with the required low value of $\OmegaM$, the fact that
approximately 1\% of the photons remain after a full period for
our assumed loss, and an assumed detection efficiency of 70 \%,
this implies a detection rate of approximately 100 photons per
hour in the revival interval. This means that a measurement time
of order 30 minutes should give convincing statistics.

After every single run of the experiment the mirror has to be
damped to reset it to its initial (thermal) state. This could be
done by electric or magnetic fields, e.g. following Ref.
\cite{Wago1998}, where a Nickel sphere was attached to the
cantilever, whose $Q$ could then be changed by 3 orders of
magnitude by applying a magnetic field.

Since the width of the revival peak scales like $1/\sqrt{T}$, the
required measurement time can also be decreased by decreasing the
temperature below 60 $\mu$K. Passive cooling techniques may be
improved. In addition, active and passive optical cooling of mirror oscillators has
been proposed \cite{Mancini1998}, and implemented
experimentally for a large mirror \cite{Cohadon1999} and for small mirrors \cite{Kleckner2006a, Arcizet2006, Gigan2006, Kippenberg2008}. Ground state cooling of the center of mass motion is achievable and reduces the required measurement time, and thus the stability requirements, by a factor of approximately 50.

%% file: Part6_Genes.tex
In this last section we turn our attention to non-classical states involving both, light and mechanics. In our discussion these will be primarily entangled states, which are generated either in steady state under a continuous drive field, or in a regime involving short pulses of light.

\subsection{Light mirror entanglement in steady state}\label{nonclassical:entangled:genes}

Entanglement of a mechanical oscillator with light has been predicted in a number of theoretical studies \cite{Genes2008a,paternostro_creating_2007,miao_universal_2010,vitali_optomechanical_2007,galve_bringing_2010,genes_simultaneous_2008,vitali_entangling_2007,ghobadi_optomechanical_2011,ghobadi_quantum_2011,abdi_effect_2011} and would be an intriguing demonstration of optomechanics in the quantum regime. These studies, as well as similar ones investigating entanglement among several mechanical oscillators \cite{mancini_entangling_2002,zhang_quantum-state_2003,pinard_entangling_2005,pirandola_macroscopic_2006,hartmann_steady_2008,vacanti_optomechanical_2008,huang_entangling_2009,vitali_stationary_2007,ludwig_entanglement_2010}, explore entanglement in the \emph{steady-state regime}. In this regime the optomechanical system is driven by one or more continuous-wave light fields and settles into a stationary state, for which the interplay of optomechanical coupling, cavity decay, damping of the mechanical oscillator, and thermal noise forces may remarkably give rise to persistent entanglement between the intracavity field and the mechanical oscillator.


The simplest example of such a scheme involves an optomechanical system driven by one continuous wave laser field. To identify conditions for good
optomechanical generation of entanglement we answer a first question that concerns
the optimal detuning of the driving laser with respect to the cavity
field. Given the form of the linearized radiation pressure
Hamiltonian $\hbar \gom (\ah+\ah^{\dag})(\bh+\bh^{\dag})$, and the
time evolution of operators with frequencies $\Delta$ and $\OmegaM$,
we focus on resonant processes where $\Delta=\pm\OmegaM$. The first
case we analyze is blue-detuning $\Delta=-\OmegaM$ and in which we split the interaction in two kind of
interactions well known in quantum optics: i) beam splitter
interaction $\hbar \gom (\ah^{\dag}\bh+h.c.)$ and ii)
down-conversion interaction, $\hbar \gom (\ah\bh+h.c.)$. Since the
beam splitter term is off-resonant by $2\OmegaM$ and also cannnot
produce entanglement starting from classical states we drop it and
focus on the down-conversion term, known to produce bipartite
entanglement. Following a standard treatment to obtain the
covariance matrix in steady state (even analytically for this
particular case), it can be shown that the logarithmic negativity
scales up with $\gom$ as
\begin{equation}
E_{\mathcal{N}}\leq \ln \left[ \frac{1+\gom/\sqrt{2\kappa
\GammaM}}{1+\nth}\right].
\end{equation}

\begin{figure}[t]
\includegraphics[width=0.99\columnwidth]{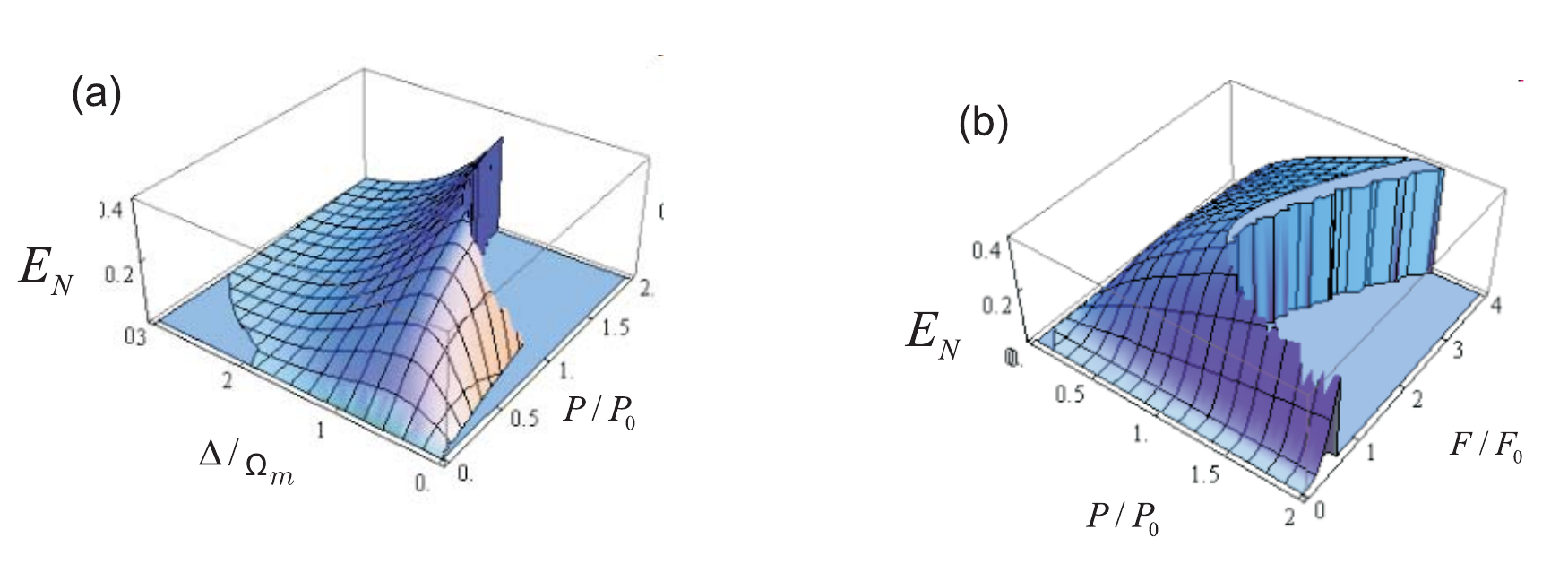}
\caption {(a) Logarithmic negativity $E_{\mathcal{N}}$ versus the
normalized detuning $\Delta/\OmegaM$ and normalized input power
$\Pin/\Pin^0$, ($\Pin^0=50$ mW) at a fixed value of the cavity
finesse $\finesse=\finesse_0=1.67 \times 10^4$; (b)
$E_{\mathcal{N}}$ versus the normalized finesse
$\finesse/\finesse_0$ and normalized input power $\Pin/\Pin^0$ at a
fixed detuning $\Delta =
\OmegaM$. Parameter values are $\OmegaM/2\pi =10$ MHz, $%
\mathcal{Q}=10^5$, mass $\mass=10$ ng, a cavity of length
$\cavlength=1$ mm driven by a laser with wavelength $810$ nm,
yielding $\gom=0.95$ KHz and a cavity bandwidth $\kappa=0.9 \OmegaM$
when $\finesse=\finesse_0$. We have assumed a reservoir temperature
for the mirror $T=0.4$ K, corresponding to $\nth\simeq 833$. The
sudden drop to zero of $E_{\mathcal{N}}$ corresponds to entering the
instability region.} \label{nonclassical:genes:fig:3}
\end{figure}

However, the system is unstable in the "blue-detuned" regime owing
to the fast transfer of energy from the cavity field to the mirror
and an unavoidable bound is found $\gom<\sqrt(2\kappa\GammaM)$ which
in consequence limits $E_{\mathcal{N}}\leq \ln 2$. Moreover, any
thermal quantum $\nth>0$ completely destroys the entanglement. One
therefore concludes that the choice of the practical operation
regime is dictated by the stability of the system. Thus, we move
into the "red detuned" which allows for larger $\gom$ by paying the
price that, for example at $\Delta=\OmegaM$ the down-conversíon
process is $2\OmegaM$ off-resonant. In this regime analytical
results are possible but cumbersome and we settled for numerically
showing the behavior of $E_{\mathcal{N}}$ in Fig.
\ref{nonclassical:genes:fig:3}a as it scales with increasing input
power and varying detunings, and in \ref{nonclassical:genes:fig:3}b
with input power and cavity finesse.

Having concluded that intracavity optomechanical entanglement is
attainable, the question of detection is to be answered next. As
detailed in \cite{vitali_optomechanical_2007} a simple scheme can be
conceived that consists of a second cavity adjacent to the main one;
the second cavity output, when weakly driven, does not modify much
the first cavity dynamics and its output light gives a direct
measurement of the mirror dynamics. With homodyne detection of both
cavities and manipulation of the two local oscillators phases one
can determine all of the entries of the covariance matrix and them
numerically extract the logarithmic negativity.

\begin{figure}[t]
\includegraphics[width=0.99\columnwidth]{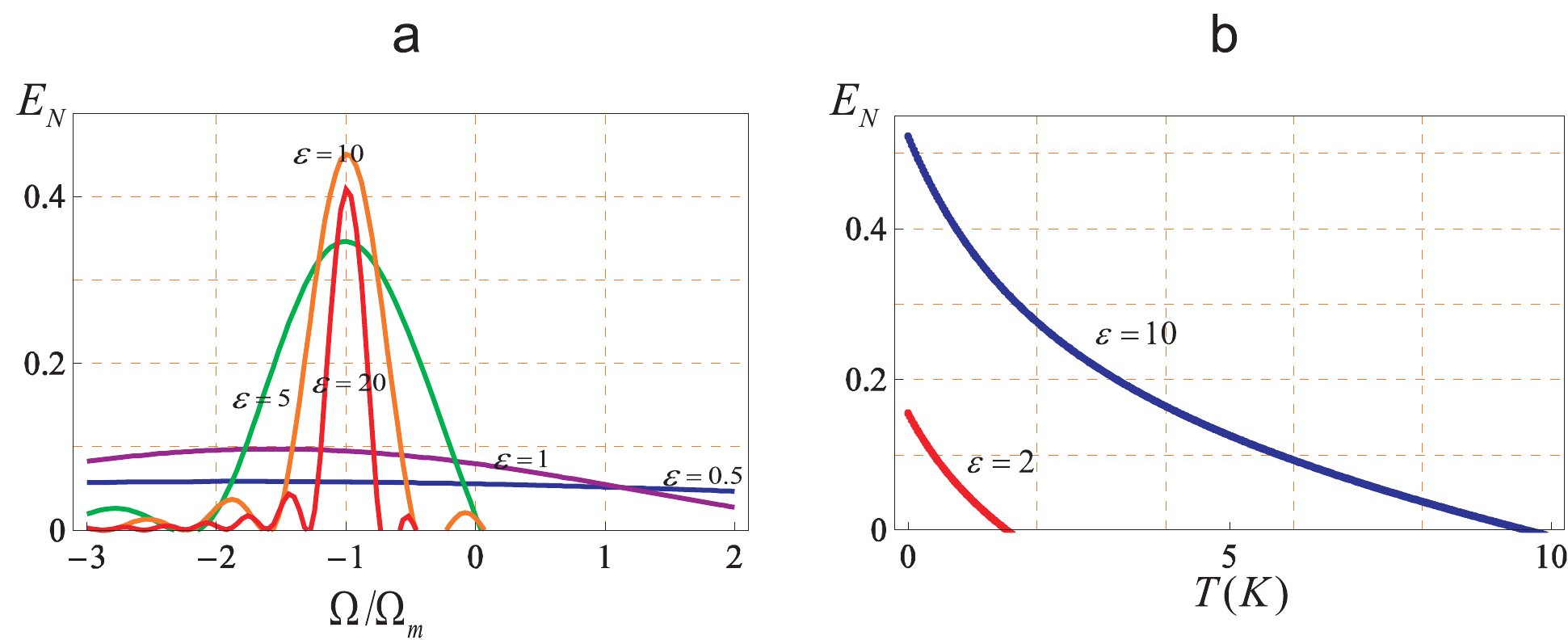}
\caption {(a) Logarithmic negativity $E_{\mathcal{N}}$ vs.
$\Omega/\OmegaM$ for different values of $\epsilon$. Optimal
entanglement of the Stokes sideband with the mirror can be deduced.
(b) $E_{\mathcal{N}}$ of output Stokes mode with mirror versus $T$
for two different values of its inverse bandwidth $\epsilon=2,10$.}
\label{nonclassical:genes:fig:4}
\end{figure}

As previously mentioned, in the "red-detuning" regime, the
down-conversion process is off-resonant and its effect much washed
out by the presence of the stronger beam-splitter interaction.
However, proper detection around the Stokes sideband (which carries
the photons entangled with the mirror) in the sense described in
\ref{nonclassical:light:genes} can extract optimal light-mirror
entanglement \cite{Genes2008a}. We show this by
choosing a central frequency of the detected mode $\Omega$ and its
bandwidth $\tau^{-1}$ and computing the bipartite system negativity.
The results are shown in Fig. \ref{nonclassical:genes:fig:4}a, where
$E_{\mathcal{N}}$ is plotted versus $\Omega/\OmegaM$ at five
different values of $\epsilon=\tau \OmegaM$ and the other parameters
similar to the ones used for Fig. \ref{nonclassical:genes:fig:3}. If
$\epsilon<1$, i.e., the bandwidth of the detected mode is larger
than $\OmegaM$, the detector does not resolve the motional
sidebands, and $E_{\mathcal{N}}$ has a value (roughly equal to that
of the intracavity case) which does not essentially depend upon the
central frequency. For smaller bandwidths (larger $\epsilon$), the
sidebands are resolved by the detection and the role of the central
frequency becomes important. In particular $E_{\mathcal{N}}$ becomes
highly peaked around the Stokes sideband $\Omega=-\OmegaM$, showing
that the optomechanical entanglement generated within the cavity is
mostly carried by this lower frequency sideband. What is relevant is
that the optomechanical entanglement of the output mode is
significantly larger than its intracavity counterpart and achieves
its maximum value at the optimal value $\epsilon \simeq 10$, i.e., a
detection bandwidth $\tau^{-1} \simeq \OmegaM/10$. This means that
in practice, by appropriately filtering the output light, one
realizes an effective entanglement distillation because the selected
output mode is more entangled with the mechanical resonator than the
intracavity field.

It is finally important to see what the robustness of the
entanglement is with increasing temperature of the thermal
reservoir. This is shown by Fig. \ref{nonclassical:genes:fig:4}b,
where the entanglement $E_{\mathcal{N}}$ of the output mode centered
at the Stokes sideband is plotted versus the temperature of the
reservoir at two different values of the bandwidth, the optimal one
$\epsilon=10$, and at a larger bandwidth $\epsilon =2$. We see the
expected decay of $E_{\mathcal{N}}$ for increasing temperature, but
above all that also this output optomechanical entanglement is
robust against temperature because it persists even above liquid He
temperatures, at least in the case of the optimal detection
bandwidth $\varepsilon=10$.

%% file: Part7_Hammerer.tex
An alternative approach to achieving optomechanical entanglement works in a \emph{pulsed regime}. It does not rely on the existence of a stable steady state, such that entanglement is not limited by stability requirements. In fact it is possible to operate in a parameter regime where a stationary state does not exist. This sort of optomechanical entanglement can be verified by using a pump--probe sequence of light pulses. The quantum state created in this protocol exhibits Einstein--Podolsky--Rosen (EPR) type entanglement \cite{Nonclassical:Einstein1935} between the mechanical oscillator and the light pulse. It thus provides the canonical resource for quantum information protocols involving continuous variable (CV) systems \cite{braunstein_quantum_2005}. Optomechanical EPR entanglement can therefore be used for the teleportation of the state of a propagating light pulse onto a mechanical oscillator as suggested in \cite{mancini_scheme_2003,pirandola_continuous-variable_2003}.

  \begin{figure}
  \begin{center}
    \includegraphics{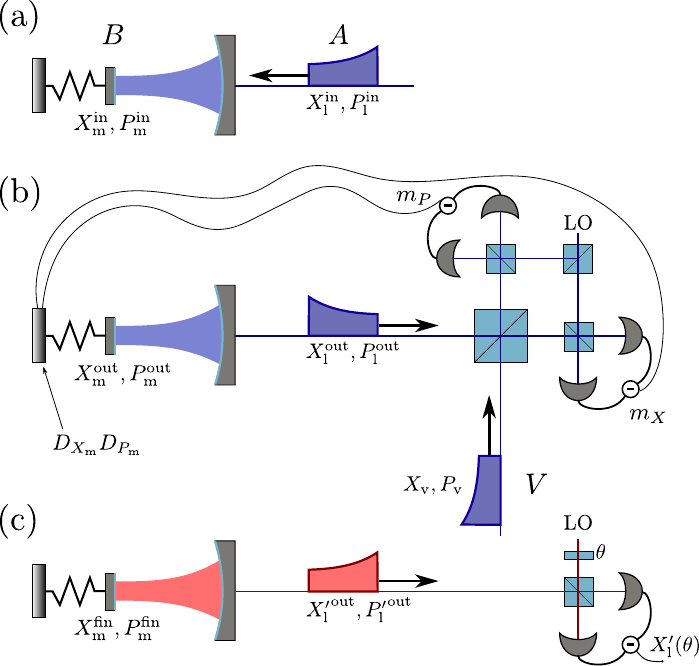}
    \caption{Schematic of the system and the proposed teleportation protocol: (a) A blue detuned light pulse (A) is entangled with the mirror (B). (b) A second light pulse (V) is prepared in the input state and interferes with A on a beam-splitter. Two homodyne detectors measure $P^{\mathrm{out}}_{\mathrm{l}}+X_{\mathrm{v}}$ and $X^{\mathrm{out}}_{\mathrm{l}}+P_{\mathrm{v}}$, yielding outcomes $m_{X}$ and $m_{P}$ respectively. Feedback is applied by displacing the mirror state in phase space by a unitary transformation $D_{X_{\mathrm{m}}}\!(m_X)\,D_{P_{\mathrm{m}}}\!(m_P)$. (c) To verify the success of the protocol, the mirror state is coherently transferred to a red detuned laser pulse and a generalised quadrature $X'_{\mathrm{l}}(\theta) ={X'_{\mathrm{l}}}^{\mathrm{out}} \cos \theta + {P_{\mathrm{l}}'}^{\mathrm{out}} \sin \theta$ is measured. Repeating steps (a)--(c) for the same input state but for different phases $\theta$ yields a reconstruction of the mirror's quantum state.}
    \label{fig:protocol}
    \end{center}
  \end{figure}
Let us consider an optomechanical cavity in a Fabry--Pérot type setup, \textit{cf.}{} Fig.~\ref{fig:protocol}). A light pulse of duration $\tau$ and carrier frequency $\omega_{\mathrm{l}}$ impinges on the cavity and interacts with the oscillatory mirror mode via radiation pressure. In a frame rotating with the laser frequency, the system is described by the (effective) Hamiltonian \cite{Mancini1994}
\begin{equation}
\label{eq:1}
H=\OmegaM^{} \bh^\dagger\bh + \Delta^{} \ah^\dagger\ah + \gom \left( \bh + \bh^\dagger \right)\left( \ah + \ah^\dagger \right)
\end{equation}
where $\Delta=\omegaopt-\omega_{\mathrm{l}}$ is the detuning of the laser drive with respect to the cavity resonance. We assume the pulse to approximately be a flat-top pulse, which has a constant amplitude for the largest part, but possesses a smooth head and tail. The coupling constant $\gom$ is then given by
\begin{equation}
\label{eq:2}
\gom=\gomzero \sqrt{\frac{2\kappa N_{\mathrm{ph}}/\tau}{(\Delta^2+\kappa^2)}},
\end{equation}
with $N_{\mathrm{ph}}$ the number of photons in the pulse. It is possible to make  either the a beam-splitter like interaction ($\ah\bh^\dagger+\bh\ah^\dagger$) or the two-mode-squeezing interaction ($\ah\bh+\ah^\dagger\bh^\dagger$) resonant by tuning the laser to one of the motional sidebands $\omegaopt\pm \OmegaM$, where the blue (anti-Stokes) sideband ($\omega_{\mathrm{l}}=\omegaopt + \OmegaM$) enhances down-conversion, while the red (Stokes) sideband ($\omega_{\mathrm{l}}=\omegaopt - \OmegaM$) enhances the beam-splitter interaction \cite{Aspelmeyer2010}.
In the proposed protocol we make use of both dynamics separately: Pulses tuned to the blue sideband are applied to create entanglement, while pulses on the red sideband are later used to read out the final mirror state. A similar separation of Stokes and anti-Stokes sideband was suggested in \cite{mancini_scheme_2003,pirandola_continuous-variable_2003} by selecting different angles of reflection of a light pulse scattered from a vibrating mirror in free space.

The full system dynamics, including the dissipative coupling of the mirror and the cavity decay, are described by quantum Langevin equations \cite{gardiner_quantum_2004}, which determine the time evolution of the corresponding operators $x_{\mathrm{m}}=(\bh+\bh^\dagger)/\sqrt{2}$, $p_{\mathrm{m}}=-i(\bh-\bh^\dagger)/\sqrt{2}$ and $\ah,\ah^\dagger$. They read
\begin{eqnarray}\label{eq:3}
\dot{x}_\mathrm{m}&=&\phantom{-}\OmegaM p_{\mathrm{m}},\\
\dot{p}_\mathrm{m}&=&-\OmegaM x_{\mathrm{m}} - \GammaM\, p_{\mathrm{m}}- \sqrt{2}\, \gom \left( \ah + \ah^\dagger \right) - \sqrt{2\GammaM}\, f,\\
\dot{a}_\mathrm{c}&=&-(i\Delta^{}+\kappa) \ah - i \sqrt{2}\, \gom\, x_{\mathrm{m}} - \sqrt{2\kappa}\, a_{in},
\end{eqnarray}
where we introduced the (self-adjoint) Brownian stochastic force $f$, and quantum noise $a_{in}$ entering the cavity from the electromagnetic environment. Both $a_{in}$ and---in the high-temperature limit---$f$ are assumed to be Markovian. Their correlation functions are thus given by $\langle{a_{in}(t) a_{in}^\dagger(t')}\rangle=\delta(t-t')$ (in the optical vacuum state) and $\langle f(t)f(t')+f(t')f(t)\rangle=(2\nth+1)\delta(t-t')$ (in a thermal state of the mechanics) \cite{gardiner_quantum_2004}.

We impose the following conditions on the system's parameters. Firstly, we drive the cavity on the blue sideband ($\Delta=-\OmegaM$) and assume to work in the resolved-sideband regime ($\kappa\ll\OmegaM$) to enhance the down-conversion dynamics. Note that in this regime a stable steady state only exists for very weak optomechanical coupling \cite{ludwig_optomechanical_2008}, which poses a fundamental limit to the amount of entanglement that can be created in a continuous-wave scheme \cite{Genes2008a}. In contrast, a pulsed scheme does not suffer from these instability issues. In fact, it is easy to check by integrating the full dynamics up to time $\tau$,
that working in this particular regime yields maximal entanglement, which increases with increasing sideband-resolution $\OmegaM/\kappa$. Secondly we assume a weak optomechanical coupling $\gom\ll \kappa$, such that only first-order interactions of photons with the mechanics contribute. This minimises pulse distortion and simplifies the experimental realization of the protocol. Taken together, the conditions $\gom\ll\kappa\ll\OmegaM$ allow us to invoke the rotating-wave-approximation (RWA), which amounts to neglecting the beam-splitter term in \ref{eq:1}. Also, we neglect mechanical decoherence effects in this section. We emphasise that this approximation is justified as long as the total duration of the protocol is short compared to the effective mechanical decoherence time $1/\GammaM \nth$, where $\GammaM$ is the mechanical damping rate and $\nth$ the thermal occupation of the corresponding bath. Corrections to this simplified model will be addressed below.

Based on the assumptions above we can now simplify equations \ref{eq:3}. For convenience we go into a frame rotating with $\OmegaM$ by substituting $\ah\rightarrow \ah e^{i \OmegaM t}$, $a_{in} \rightarrow a_{in} e^{i \OmegaM t}$ and $\bh\rightarrow \bh e^{-i \OmegaM t}$. Note that in this picture the central frequency of $a_{in}$ is located at $\omega_{\mathrm{l}}-\OmegaM=\omegaopt$. In the RWA the Langevin equations then simplify to
\begin{align}\label{eq:4}
\dot{a}_\mathrm{c}&=-\kappa \ah\, - i \gom\, \bh^\dagger - \sqrt{2\kappa}\, a_{in}, &
\dot{a}_\mathrm{m}&=- i \gom\, \ah^\dagger.
\end{align}

In the limit $\gom\ll \kappa$ we can use an adiabatic solution for the cavity mode and we therefore find
\begin{align}
\ah(t) &\approx -i \frac{\gom}{\kappa} \bh^\dagger(t) - \sqrt{\frac{2}{\kappa}} a_{in}(t),\label{eq:5} \\
\bh(t) &\approx e^{G t} \bh(0) +i \sqrt{2G} e^{Gt} \intdb{0}{t}{s} e^{-Gs} a_{in}^\dagger(s),\label{eq:6}
\end{align}
where we defined $G=\gom^2/\kappa$. Equation \ref{eq:6} shows that the mirror motion gets correlated to a light mode of central frequency $\omega_\mathrm{l}-\OmegaM$ (which coincides with the cavity resonance frequency $\omegaopt$) with an exponentially shaped envelope $\alpha_{\mathrm{in}}(t) \propto e^{-Gt}$. Using the standard cavity input-output relations $a_{out}=a_{in}+\sqrt{2\kappa}\,\ah$ allows us to define a set of normalised temporal light-modes
\begin{align}\label{eq:7}
A_{\mathrm{in}}&=\sqrt{\frac{2G}{1-e^{-2G \tau}}}\intdb{0}{\tau}{t} e^{-G t}a_{in}(t), &
A_{\mathrm{out}}&=\sqrt{\frac{2G}{e^{2G\tau}-1}}\intdb{0}{\tau}{t} e^{G t}a_{out}(t),
\end{align}
which obey the canonical commutation relations $[A_{i},A^{\dagger}_{i}]=1$.
Together with the definitions $B_{\mathrm{in}}=\bh(0)$ and $B_{\mathrm{out}}=\bh(\tau)$ we arrive at the following expressions, which relate the mechanical and optical mode at the end of the pulse $t=\tau$
\begin{align}\label{eq:8}
A_{\mathrm{out}}&=-e^{G\tau} A_{\mathrm{in}} -i \sqrt{e^{2G\tau}-1}B_{\mathrm{in}}^{\dagger},&
B_{\mathrm{out}}&=e^{G\tau} B_{\mathrm{in}} + i \sqrt{e^{2G\tau}-1}A_{\mathrm{in}}^{\dagger}.
\end{align}
By expressing equations \ref{eq:8} in terms of quadratures $X_\mathrm{m}^i=(B_i+B_i^\dagger)/\sqrt{2}$ and $X_\mathrm{l}^i=(A_i+A_i^\dagger)/\sqrt{2}$, where $i \in\{\mathrm{in},\mathrm{out}\}$, and their corresponding conjugate variables, we can calculate the so-called EPR-variance $\Delta_\mathrm{EPR}$ of the state after the interaction.
For light initially in vacuum $(\Delta X_\mathrm{l}^\mathrm{in})^2=(\Delta P_\mathrm{l}^\mathrm{in})^2=\frac{1}{2}$ and the mirror in a thermal state $(\Delta X_\mathrm{m}^\mathrm{in})^2=(\Delta P_\mathrm{m}^\mathrm{in})^2=n_0+\frac{1}{2}$, the state is entangled iff \cite{duan_inseparability_2000}
\begin{eqnarray}\label{eq:11}
\Delta_\mathrm{EPR}&=&\left[\Delta(X^{\mathrm{out}}_{\mathrm{m}}+P^{\mathrm{out}}_{\mathrm{l}})\right]^2+\left[\Delta(P^{\mathrm{out}}_{\mathrm{m}}+X^{\mathrm{out}}_{\mathrm{l}})\right]^2\\
&=&2(n_0+1)\left(e^r-\sqrt{e^{2r}-1}\right)^2<2,
\end{eqnarray}
where  $r=G\tau$ is the squeezing parameter and $n_0$ the initial occupation number of the mechanical oscillator.
Note that in the limit of large squeezing $r\gg 1$ we find that the variance $\Delta_\mathrm{EPR}\approx (n_{0}+1)e^{-2r}/2$ is suppressed exponentially, which shows that the created state asymptotically approximates an EPR-state. Therefore, this state can be readily used to conduct optomechanical teleportation. Rearranging \ref{eq:11}, we find that the state is entangled as long as
\begin{equation}
\label{eq:12}
r>r_0=\frac{1}{2}\ln\left(\frac{(n_0+2)^2}{4(n_0+1)}\right) \sim \frac{1}{2}\ln{n_0},
\end{equation}
where the last step holds for $n_0\rightarrow \infty$. This illustrates that in our scheme the requirement on the strength of the effective optomechanical interaction, as quantified by the parameter $r=\frac{\gom^2 \tau}{\kappa}$, scales logarithmically with the initial occupation number $n_0$ of the mechanical oscillator. This tremendously eases the protocol's experimental realization, as neither $\gom$ nor $\tau$ can be arbitrarily increased. Note that $n_0$ need not be equal to the mean bath occupation $\nth$, but may be decreased by laser pre-cooling to improve the protocol's performance.

To verify the successful creation of entanglement a red detuned laser pulse ($\Delta=\OmegaM$) is sent to the cavity where it resonantly drives the beam-splitter interaction, and hence generates a state-swap between the mechanical and the optical mode. It is straightforward to show that choosing $\Delta=\OmegaM$ leads to a different set of Langevin equations which can be obtained from \ref{eq:4} by dropping the Hermitian conjugation ($\dagger$) on the right-hand-side. By defining modified mode functions $\alpha'_{\mathrm{in(out)}}=\alpha_{\mathrm{out(in)}}$ and corresponding light modes $A'_{\mathrm{in(out)}}$ one obtains input/output expressions in analogy to \eqref{eq:8}
\begin{align}\label{eq:13}
A'_{\mathrm{out}}&=-e^{-G\tau} A'_{\mathrm{in}} +i \sqrt{1-e^{-2G\tau}}B_{\mathrm{in}}, &
B_{\mathrm{out}}&=e^{-G\tau} B_{\mathrm{in}} - i
\sqrt{1-e^{-2G\tau}}A'_{\mathrm{in}}.
\end{align}
The pulsed state-swapping operation therefore also features an exponential scaling with $G\tau$. For $G\tau \rightarrow \infty$ the expressions above reduce to $A'_{\mathrm{out}}=-i B_{\mathrm{in}}$ and $B_{\mathrm{out}}=i A'_{\mathrm{in}}$, which shows that in this case the mechanical state---apart from a phase shift---is perfectly transferred to the optical mode. In the Schrödinger-picture this amounts to the transformation $|\varphi\rangle_{\mathrm{m}}|\psi\rangle_{\mathrm{l}}\rightarrow |\psi\rangle_{\mathrm{m}}|\varphi\rangle_{\mathrm{l}}$, where $\varphi$ and $\psi$ constitute the initial state of the mechanics and the light pulse respectively. The state-swap operation thus allows us to access mechanical quadratures by measuring quadratures of the light and therefore to reconstruct the state of the bipartite system via optical homodyne tomography.

As we have shown above, pulsed operation allows us to create EPR-type entanglement, which forms the central entanglement resource of many quantum information processing protocols \cite{braunstein_quantum_2005}. An immediate extension of the proposed scheme is an optomechanical continuous variables quantum teleportation protocol. The main idea of quantum state teleportation in this context is to transfer an arbitrary quantum state $|\psi_{\mathrm{in}}\rangle$ of a travelling wave light pulse onto the mechanical resonator, without any direct interaction between the two systems, but by making use of optomechanical entanglement. The scheme works in full analogy to the CV teleportation protocol for photons \cite{vaidman_teleportation_1994,braunstein_teleportation_1998} and, due to its pulsed nature, closely resembles the scheme used in atomic ensembles \cite{hammerer_teleportation_2005,sherson_quantum_2006}:
A light pulse (A) is sent to the optomechanical cavity and is entangled with its mechanical mode (B) via the dynamics described above. Meanwhile a second pulse (V) is prepared in the state $|\psi_{\mathrm{in}}\rangle$, which is to be teleported. This pulse then interferes with A on a beam-splitter. In the output ports of the beam-splitter, two homodyne detectors measure two joint quadratures $P^{\mathrm{out}}_{\mathrm{l}}+X_{\mathrm{v}}$ and $X^{\mathrm{out}}_{\mathrm{l}}+P_{\mathrm{v}}$, yielding outcomes $m_{X}$ and $m_{P}$ respectively. This constitutes the analogue to the Bell-measurement in the case of qubit teleportation and effectively projects previously unrelated systems A and V onto an EPR-state \cite{bouwmeester_experimental_1997}. Note that both the second pulse and the local oscillator for the homodyne measurements must be mode-matched to A after the interaction, \textit{i.e.}{}, they must possess the identical carrier frequency as well as the same exponential envelope. The protocol is concluded by displacing the mirror in position and momentum by $m_{\mathrm{X}}$ and $m_{\mathrm{P}}$ according to the outcome of the Bell-measurement. This can be achieved by means of short light-pulses, applying the methods described in \cite{Vanner2011,cerrillo_pulsed_2011}. After the feedback the mirror is then described by \cite{braunstein_quantum_2005}
\begin{eqnarray}
X_{\mathrm{m}}^{\mathrm{fin}}&=X_{\mathrm{m}}^{\mathrm{out}}+P^{\mathrm{out}}_{\mathrm{l}}+X_{\mathrm{v}}=
X_{\mathrm{v}}+\left(e^r-\sqrt{e^{2r}-1}\right)(X_{\mathrm{m}}^{\mathrm{in}}-P^{\mathrm{in}}_{\mathrm{l}}),\label{eq:15}\\
P_{\mathrm{m}}^{\mathrm{fin}}&=P_{\mathrm{m}}^{\mathrm{out}}+X^{\mathrm{out}}_{\mathrm{l}}+P_{\mathrm{v}}
=P_{\mathrm{v}}+\left( e^r-\sqrt{e^{2r}-1} \right)(P_{\mathrm{m}}^{\mathrm{in}}-X^{\mathrm{in}}_{\mathrm{l}}),\label{eq:16}
\end{eqnarray}
which shows that its final state corresponds to the input state plus quantum noise contributions. It is obvious from these expressions that the total noise added to both quadratures (second term in \ref{eq:15} and \ref{eq:16} respectively) is equal to the EPR-variance. Again, for large squeezing $r\gg 1$ the noise terms are exponentially suppressed and in the limit $r\rightarrow \infty$, where the resource state approaches the EPR-state, we obtain perfect teleportation fidelity, \textit{i.e.}{}, $X^{\mathrm{fin}}_{\mathrm{m}}=X_{\mathrm{v}}$ and $P^{\mathrm{fin}}_{\mathrm{m}}=P_{\mathrm{v}}$. In particular this operator identity means, that \emph{all} moments of $X_{\mathrm{v}}$, $P_{\mathrm{v}}$ with respect to the input state $|\psi_{\mathrm{in}}\rangle$ will be transferred to the mechanical oscillator, and hence its final state will be identically given by $|\psi_{\mathrm{in}}\rangle$.

We found that in the ideal scenario the amount of entanglement essentially depends only on the coupling strength (or equivalently on the input laser power) and the duration of the laser pulse and that it shows an encouraging scaling, growing exponentially with $G\tau$. This in turn means that the minimal amount of squeezing needed to generate entanglement only grows logarithmically with the initial mechanical occupation $n_0$. In a more realistic scenario one has to include thermal noise effects and effects of counter-rotating terms. Including the above-mentioned perturbations results in a final state which deviates from an EPR-entangled state. To minimise the extent of these deviations, the system parameters must obey the following conditions:

\begin{enumerate}
\item $\kappa\ll\OmegaM$ results in a sharply peaked cavity response and implies that the down-conversion dynamics is heavily enhanced with respect to the suppressed beam-splitter interaction.
\item $\gom < \kappa$ inhibits multiple interactions of a single photon with the mechanical mode before it leaves the cavity. This suppresses spurious correlations to the intracavity field. It also minimises pulse distortion and simplifies the protocol with regard to mode matching and detection.
\item $\gom \tau \gg 1$ is needed in order to create sufficiently strong entanglement. This is due to the fact that the squeezing parameter $r=(\gom/\kappa) \gom\tau$ should be large, while $\gom/\kappa$ needs to be small.
\item $\nth \GammaM \tau\ll 1$, where $\nth$ is the thermal occupation of the mechanical bath, assures coherent dynamics over the full duration of the protocol, which is an essential requirement for observing quantum effects. As the thermal occupation of the mechanical bath may be considerably large even at cryogenic temperatures, this poses (for fixed $\GammaM$ and $\nth$) a very strict upper limit to the pulse duration $\tau$.
\end{enumerate}

Note however that not all of these inequalities have to be fulfilled equally strictly, but there rather exists an optimum which arises from balancing all contributions. It turns out that fulfilling (4) is critical for successful teleportation, whereas (1)--(3) only need to be weakly satisfied. Taking the above considerations into account, we find a sequence of parameter inequalities
\begin{equation}
\label{eq:18}
\nth\GammaM \ll \frac{1}{\tau}\ll \gom \ll \kappa \ll \OmegaM,
\end{equation}
which defines the optimal parameter regime. Dividing this equation by $\GammaM$ and taking a look at the outermost condition $\nth \ll Q$, where $Q=\OmegaM/\GammaM$ is the mechanical quality factor, we see that the ratio $Q/\nth$ defines the range which all the other parameters have to fit into. It is intuitively clear, that a high quality factor and a low bath occupation number, and consequently low effective mechanical decoherence, are favourable for the success of the protocol. Equivalently, we can rewrite the occupation number as $\nth=k_{\mathrm{B}}T_{\mathrm{bath}}/\hbar \OmegaM$ and therefore find $k_{\mathrm{B}}T_{\mathrm{bath}}/\hbar \ll Q\cdot\OmegaM$, where now the $Q\cdot f$-product ($f=\OmegaM/2\pi$) has to be compared to the thermal frequency of the bath.
Let us consider a numerical example: For a temperature $T_{\mathrm{bath}}\approx 100\,\mathrm{mK}$ the left-hand-side gives $k_{\mathrm{B}}T_{\mathrm{bath}}/\hbar\approx 2\pi\cdot 10^{9}\,\mathrm{Hz}$. The $Q \cdot f$-product consequently has to be several orders of magnitude larger to successfully create entanglement. As current optomechanical systems feature a $Q\cdot f$-product of $2\pi \cdot 10^{11}\,\mathrm{Hz}$ and above \cite{cole_phonon-tunnelling_2011,ding_high_2010,eichenfield_optomechanical_2009,safavi-naeini_electromagnetically_2011}, this requirement seems feasible to meet.

%% file: Part8_Hammerer.tex
The selection of protocols presented in the present chapter clearly demonstrate the feasibility --- and the stringent requirements --- for quantum state engineering in mesoscopic mechanical systems. At the time of writing this book chapter state of the art experiments achieve ground state cooling, but manifest quantum effects in the dynamics of optomechanical systems have yet to be demonstrated. From the discussion given above it should be clear that a necessary condition for observing quantum effects is a sufficiently large product of mechanical quality factor and frequency, $Q\OmegaM>k_BT/\hbar$. What exactly ``large'' means in this context depends crucially on the protocol to be implemented. The goal of making the quantum regime accessible for mechanical systems thus has to  be approached from both sides: Experimentally, by developing optomechanical systems with a sufficiently large  $Q\OmegaM$-product; and theoretically, by developing schemes which are not too demanding regarding the magnitude of this number. Once those two ends meet this will mark the birth of a new field of research, quantum optomechanics. 